\documentclass[useAMS,usenatbib]{mn2e}

\usepackage{verbatim} 
\usepackage{natbib} 
\usepackage{amsmath} 
\usepackage{amsbsy}
\usepackage{amssymb}
\usepackage{mathrsfs} 
\usepackage{lscape} 
\usepackage{graphicx}
\usepackage{epstopdf}
\usepackage{deluxetable}
\usepackage{ctable} 
\usepackage{fixltx2e}

\title[UVB Model]{A Cosmic UV/X-ray Background Model Update}
\author[Faucher-Gigu\`ere]{Claude-Andr\'e Faucher-Gigu\`ere\thanks{cgiguere@northwestern.edu}\\
Department of Physics and Astronomy and Center for Interdisciplinary Exploration and Research in Astrophysics (CIERA),\\ Northwestern University, 2145 Sheridan Road, Evanston, IL 60208, USA 
}

\begin{document}
\maketitle

\begin{abstract}
We present an updated model of the cosmic ionizing background from the UV to the X-rays. 
Relative to our previous model (Faucher-Gigu\`ere et al. 2009), the new model provides a better match to a large number of up-to-date empirical constraints, including: 1) new galaxy and AGN luminosity functions; 2) stellar spectra including binary stars; 
3) obscured and unobscured AGN; 4) a measurement of the non-ionizing UV background; 5) measurements of the intergalactic HI and HeII photoionization rates at $z\sim 0 - 6$; 6) the local X-ray background; and 7) improved measurements of the intergalactic opacity. 
In this model, AGN dominate the HI ionizing background at $z \lesssim 3$ and star-forming galaxies dominate it at higher redshifts. 
Combined with the steeply declining AGN luminosity function beyond $z\sim2$, the slow evolution of the HI ionization rate inferred from the high-redshift HI Ly$\alpha$ forest requires an escape fraction from star-forming galaxies that increases with redshift (a population-averaged escape fraction of $\approx1\%$ suffices to ionize the intergalactic medium at $z=3$ when including the contribution from AGN). 
We provide effective photoionization and photoheating rates calibrated to match the Planck 2018 reionization optical depth and recent constraints from the HeII Ly$\alpha$ forest in hydrodynamic simulations.
\end{abstract}

\begin{keywords}
Cosmology: theory, diffuse radiation, reionization --- galaxies: formation, active, intergalactic medium
\vspace{-0.5cm}
\end{keywords}

\section{Introduction}
\label{sec:intro}
Galaxies and active galactic nuclei (AGN) produce a diffuse background of ultra-violet (UV) radiation that permeates the intergalactic medium (IGM). 
This cosmic UV background (UVB) keeps the IGM ionized following the epoch of reionization. 
UVB models are widely used for two purposes. 
The first is ionization corrections, which are necessary to convert measured abundances of ions to total masses in different elements \citep[e.g.,][]{2016ApJ...830...87S, 2017ApJ...837..169P, 2017ApJ...842L..19C}. 
The second is to model the ionization and cooling of cosmic gas in cosmological hydrodynamic simulations \citep[e.g.,][]{1992ApJ...393...22C, 1996ApJS..105...19K, 2009MNRAS.393...99W, 2012ApJS..202...13G}. 
AGN also produce a background of X-ray radiation which can ionize heavy elements and further heat the IGM through inverse Compton scattering \citep[][]{1999ApJ...517L...9M}. 
The cosmic X-ray background (CXB) is directly measured at $z=0$ by X-ray telescopes \citep[e.g.,][]{1962PhRvL...9..439G, 2007ApJ...661L.117H, 2008ApJ...689..666A}. 
For simplicity, we will generally refer to the UVB in this paper, but we also include X-rays from AGN. 

Several authors have modeled the spectrum and redshift evolution of the UVB \citep[e.g.,][]{1990ApJ...350....1M, 1996ApJ...461...20H, 1996ApJS..102..191G, 1998AJ....115.2206F, 2009ApJ...703.1416F}. 
Of these, the series of models by Haardt \& Madau \citep[][]{1996ApJ...461...20H, 2001cghr.confE..64H, 2012ApJ...746..125H} have been the most widely used. 
We refer to the \cite{2012ApJ...746..125H} model as HM12. 
In the last decade, the \citet[][hereafter FG09]{2009ApJ...703.1416F} model has also been used in a number of studies, e.g. in the Illustris and FIRE galaxy formation simulations \citep[][]{2014MNRAS.444.1518V, 2018MNRAS.473.4077P, 2014MNRAS.445..581H, 2018MNRAS.480..800H}. 
Recently, \citet[][hereafter KS19]{2019MNRAS.484.4174K} produced new synthesis models of the extragalactic background light from the far infrared to the TeV $\gamma-$rays and \citet[][hereafter P19]{2019MNRAS.485...47P} updated the HM12 model with an emphasis on the effects of reionization. 
\begin{figure*}
\begin{center}
\includegraphics[width=0.99\textwidth]{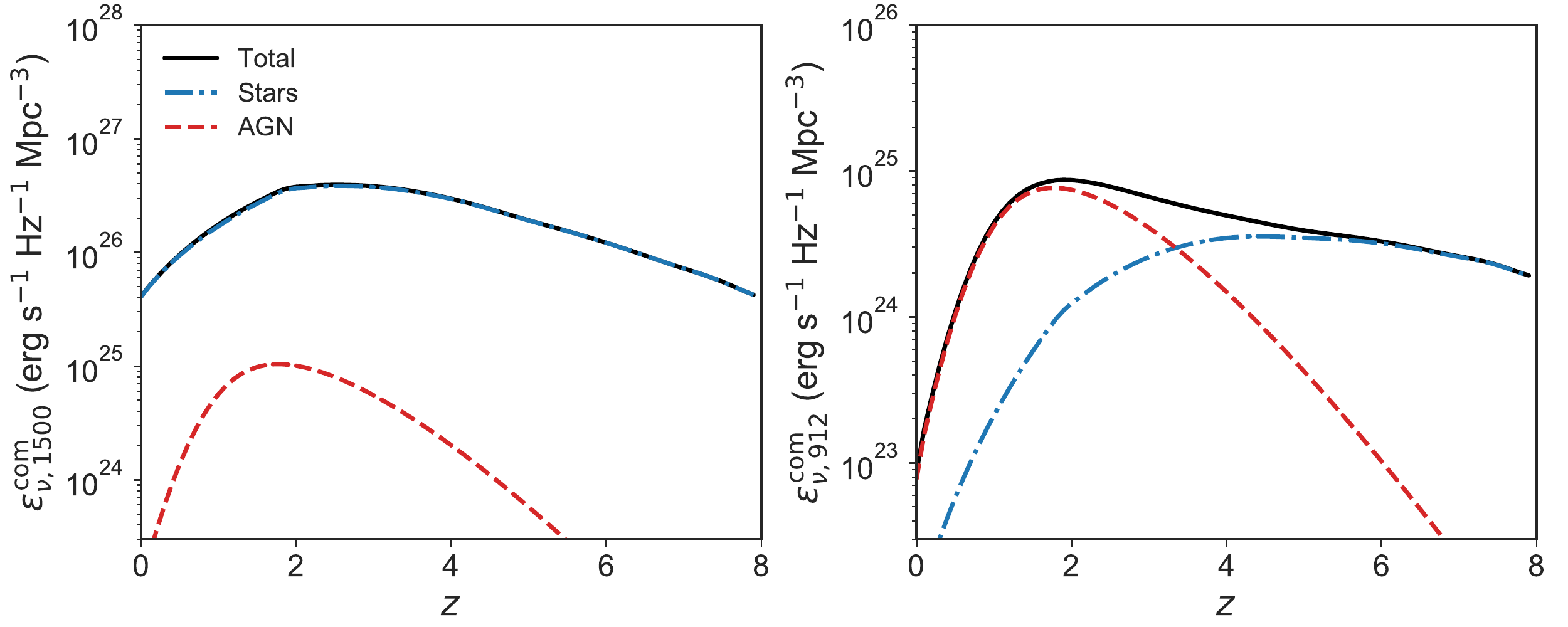}
\end{center}
\caption{Comoving emissivities as a function of redshift. 
The total (solid black) is the sum of a contribution from star-forming galaxies (dot-dashes) and AGN (dashes). 
The left panel shows the non-ionizing emissivities at rest-frame UV wavelength 1500~\AA~and the right panel shows the emissivities from the same populations just above HI photoionization edge (912~\AA). 
The different relative contributions from AGN and star-forming galaxies at 1500~\AA~and at 912~\AA~arise primarily from different assumed escape fractions (unity for AGN but increasing with redshift for galaxies; see eq. \ref{eq:fesc_stars}).}
\label{fig:eps_vs_z} 
\end{figure*}

Despite the fact that a UVB model is assumed in most cosmological hydrodynamic simulations and that many observational inferences from quasar absorption data require an ionization correction, there is a relative paucity of available models which take into account the most up-to-date empirical constraints on the background intensity and its sources. 
This not only means that widely-used models may not be consistent with all the latest observational data, but also that researchers are often unable to assess uncertainties in inferences based on these models. 
For example, a comparison of the ionization rate required by the low-redshift Ly$\alpha$ forest to the HM12 model suggested that known astrophysical sources could not explain the observationally-inferred intergalactic ionization rate, implying a ``photon underproduction crisis'' \citep[][]{2014ApJ...789L..32K}. 
However, as subsequent studies pointed out, the tension between the Ly$\alpha$ forest observations and UVB models could be alleviated by considering the FG09 model instead, since the latter predicted a higher UVB amplitude at $z\sim0-0.5$ than HM12 \citep[e.g.,][see also Khaire \& Srianand 2015]{2015ApJ...811....3S, 2017ApJ...835..175G}.\nocite{2015MNRAS.451L..30K} 
Our new UVB synthesis model, which provides a good match to the most recent measurements of the low-redshift ionization rate, confirms that there is at present no crisis, since most or all of the low-redshift ionization rate can be explained by AGN (see \S \ref{sec:integral_constraints}).

In this paper, we present an update of the FG09 UVB model. 
Our methods are similar to those described in FG09, but we incorporate several more more up-to-date empirical constraints. 
Specifically, our new model is informed by:
\begin{enumerate}
\item Recent measurements of the galaxy UV luminosity function out to the epoch of HI reionization and of the total cosmic UV emissivity at low redshifts. 
\item A stellar spectral template including binary stars, as well as recent observational constraints on dust attenuation and the escape of ionizing photons from galaxies.
\item New measurements of the AGN luminosity function.
\item An AGN spectral template including both obscured and unobscured sources and constrained to match the local X-ray background.
\item Improved measurements of the IGM opacity.
\item Updated constraints from the Ly$\alpha$ forest on the integrated HI photoionization rate, especially at low redshift ($z<0.5$) and approaching the epoch of HI reionization ($z\sim5-6$). 
\item The Planck 2018 constraint on reionization from the optical depth to the surface of last scattering.
\item The latest observational constraints on HeII reionization.
\end{enumerate}

Another important motivation for our new calculations is to provide ``effective'' UVB models designed to address an important issue with the use of spatially homogeneous UVB models in cosmological simulations. 
Namely, most current simulation codes calculate ionization states under the assumption of optically thin gas in photoionization equilibrium with an homogeneous ionizing background. 
While this approximation is valid in the IGM when the mean free path of ionizing photons is large, it breaks down before and during reionization events. 
During reionization, IGM patches are not in photoionization equilibrium but rather transition rapidly from neutral to ionized as ionization fronts propagate. 
As a result, simulations that use a homogeneous UVB model do not in general correctly model the timing of reionization events and the photoheating that accompanies reionization \citep[e.g.,][]{2015MNRAS.450.4081P}. 

This issue was briefly discussed in FG09, and was partially addressed by the 2011 release of an updated version of the FG09 UVB model modified to produce a HI reionization redshift $z_{\rm rei,HI}\sim10$ consistent with the WMAP 7-year optical depth \citep[][]{2011ApJS..192...18K}. 
This model however produces a reionization redshift substantially earlier than the best fit $z_{\rm rei,HI}=7.82\pm0.71$ from the Planck 2018 results. 
Relative to earlier Planck results, the 2018 measurement of the electron scattering optical depth to the surface of last scattering ($\tau_{\rm e}=0.054\pm0.007$) benefited from improved measurements of the large-scale polarization of the cosmic microwave background \citep[CMB;][]{Planck-Collaboration:2018aa}. 

Recently, \cite{2017ApJ...837..106O} substantially developed the concepts of effective photoionization and photoheating rates and showed explicitly how to modify UVB models so that they produce a correct mean reionization history and more accurate photoheating when used in standard simulation codes. 
\cite{2019MNRAS.485...47P} published a modified version of the HM12 UVB model incorporating similar effective rates, but calibrated to produce a higher optical depth $\tau_{\rm e} = 0.065$, closer to the Planck 2015 results \citep{2016A&A...594A..13P}.

The plan of this paper is as follows. 
We review our UVB modeling methodology, model ingredients, and compare our new spectral synthesis results to observational constraints in \S \ref{sec:models}. 
We then derive effective photoionization and photoheating rates calibrated to match desired reionization histories in \S \ref{sec:reionization}. 
We discuss our results and conclude in \S \ref{sec:discussion}. 
A series of appendices provide additional details, including on how the results depend on model assumptions.

Throughout, we assume a standard flat $\Lambda$CDM cosmology with $\Omega_{\rm m }=0.32$, $\Omega_{\Lambda}=1-\Omega_{\rm m}$, $\Omega_{\rm b}=0.049$, and $H_{0}=67$ km s$^{-1}$ Mpc$^{-1}$ \citep[e.g.,][]{Planck-Collaboration:2018aa}. 
We use $X$ to denote the the hydrogen mass fraction and $Y$ for the helium mass fraction. 
Lower-case $x$ and $y$ are used to denote ionized fractions of hydrogen and helium, respectively. 
For example, $x_{\rm II}$ is the hydrogen mass fraction in HII, and $y_{\rm II}$ and $y_{\rm III}$ are the helium mass fractions in HeII and HeIII. 
\begin{figure}
\begin{center}
\includegraphics[width=0.475\textwidth]{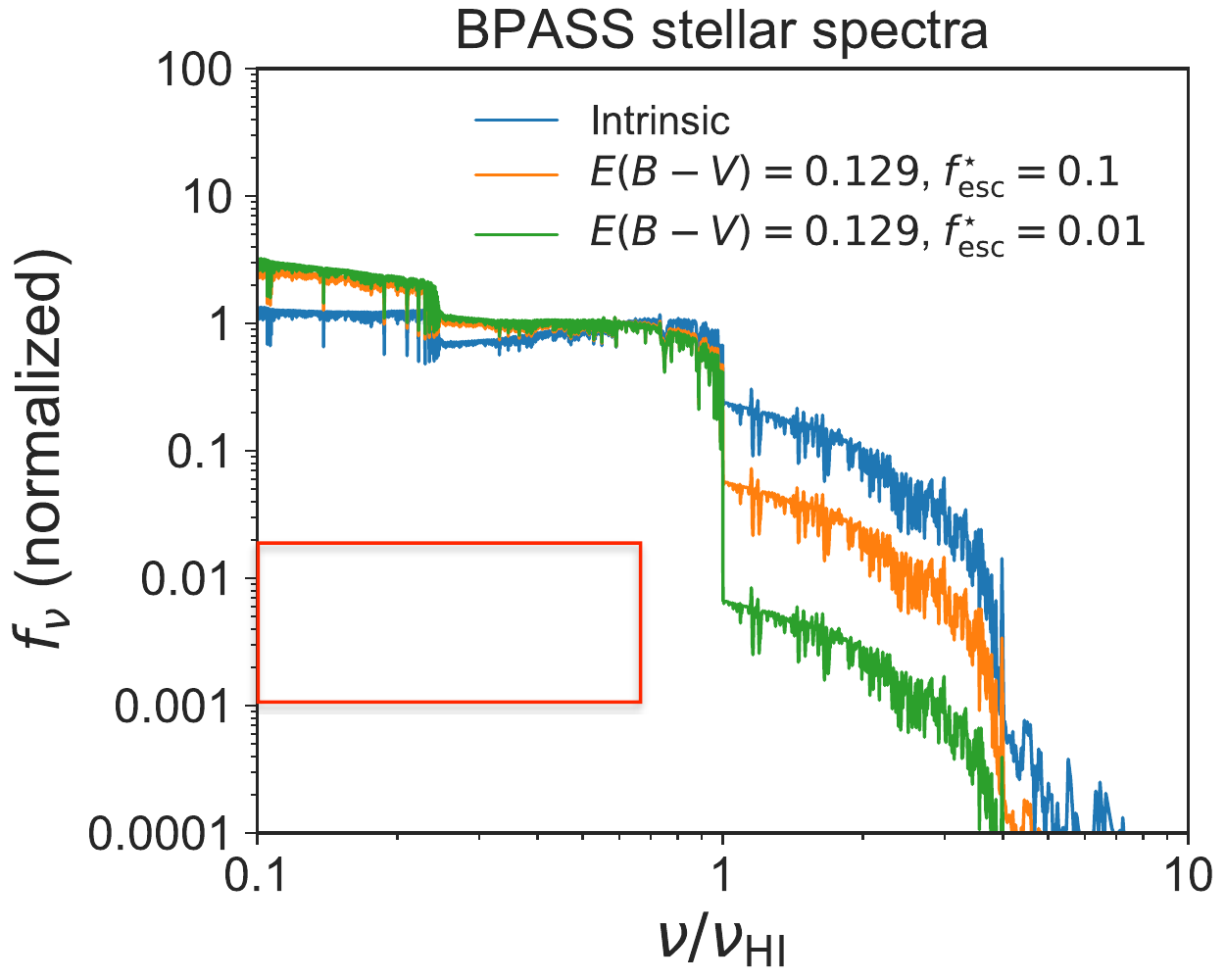}
\end{center}
\caption[]{ 
BPASS spectral templates for star-forming galaxies. 
The parameters are motivated by observations of Lyman break galaxies at $z\sim2-3.5$ (see \S \ref{sec:stellar_spec}). 
We assume a ``holes'' model for the escape of ionizing photons, i.e. that ionizing photons escape the ISM through a fraction $f_{\rm esc}^{\star}$ of clear sight lines. 
The blue curve shows the spectral template before dust attenuation and assuming an unity escape fraction (the intrinsic spectrum of the stellar population). 
The orange curve assumes an extinction $E(B-V)=0.129$ and an escape fraction $f_{\rm esc}^{\star}=0.1$. 
The green curve assumes the same extinction but an  escape fraction $f_{\rm esc}^{\star}=0.01$. 
All spectra shown here are normalized at wavelength $\lambda = 1500$~\AA.
}
\label{fig:BPASS} 
\end{figure}

\begin{figure}
\begin{center}
\includegraphics[width=0.475\textwidth]{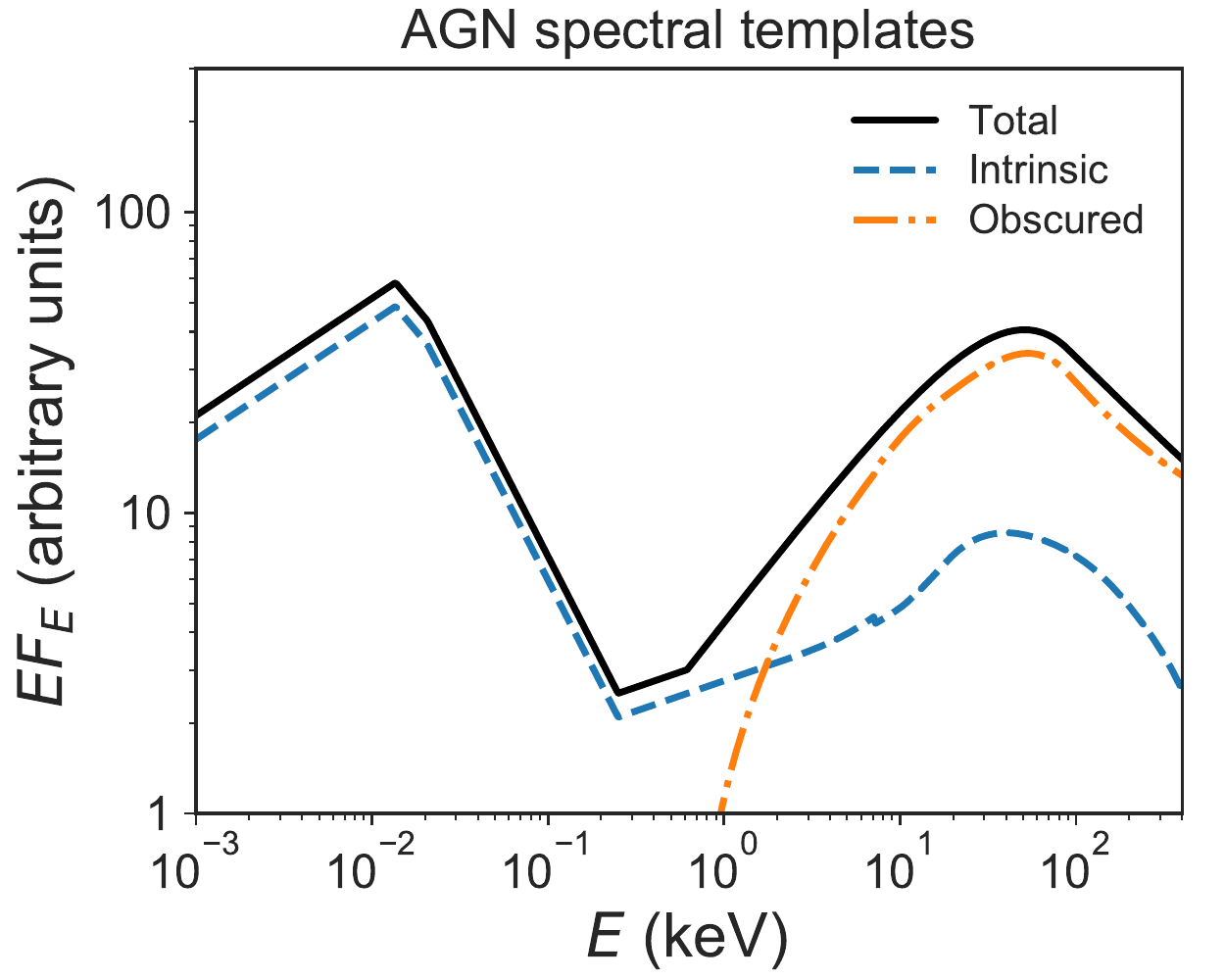}
\end{center}
\caption[]{AGN spectral templates. 
The mean intrinsic AGN spectral energy distribution (SED) is shown by the dashed curve and the inferred mean SED for obscured AGN is indicated by the dash-dotted curve (see \S \ref{sec:agn_spec} for details). 
The total AGN spectral template, averaged over both obscured and non-obscured sources, is shown by the solid curve. 
In this plot, the templates are normalized so that the intrinsic and obscured templates sum to the total for an assumed obscuration fraction $f_{\rm obsc}=0.75$. 
The total template is then offset by a factor of 1.2 for visual clarity.
}
\label{fig:AGN_spectra} 
\end{figure}

\section{UV/X-ray background modeling}
\label{sec:models}

\subsection{Radiative transfer equations}
\label{sec:rad_transp} 
The angle-averaged specific intensity of the homogeneous background is denoted by $J_{\nu}$ and satisfies the cosmological radiative transfer equation,
\begin{equation}
\label{transfer equation}
\left(
\frac{\partial}{\partial t} - \nu H \frac{\partial}{\partial \nu}
\right)
J_{\nu}
=
-3 H J_{\nu}
-c \alpha_{\nu} J_{\nu}
+ \frac{c}{4\pi}\epsilon_{\nu},
\end{equation}
where $H(t)$ is the Hubble parameter, $c$ is the speed of light, $\alpha_{\nu}$ is the proper absorption coefficient per unit length, and $\epsilon_{\nu}$ is the proper emissivity \citep[e.g.,][]{1996ApJ...461...20H}. 
We consider three components to the total emissivity, corresponding to star-forming galaxies (superscript $\star$), AGN (superscript AGN), and recombination emission (superscript rec):
\begin{equation}
\epsilon_{\nu} = \epsilon_{\nu}^{\star} + \epsilon_{\nu}^{\rm AGN} + \epsilon_{\nu}^{\rm rec}.
\end{equation}
We discuss our assumptions for the different emissivity components in \S \ref{sec:stellar_spec}-\ref{sec:recombs}.

Integrating equation (\ref{transfer equation}) and expressing the result in terms of redshift gives
\begin{equation}
\label{transfer equation solution}
J_{\nu_{0}}(z_{0})
=\frac{1}{4 \pi}
\int_{z_{0}}^{\infty}
dz
\frac{dl}{dz}
\frac{(1+z_{0})^{3}}{(1+z)^{3}} \epsilon_{\nu}(z) \exp[-\bar{\tau}(\nu_0, z_{0}, z)],
\end{equation}
where $\nu=\nu_{0}(1+z)/(1+z_{0})$, the proper line element $dl/dz=c/[(1+z)H(z)]$, and the ``effective optical depth'' $\bar{\tau}$ quantifies the attenuation due to absorption in the IGM of photons of frequency $\nu_{0}$ at redshift $z_{0}$ that were emitted at redshift $z$ through $e^{\bar{\tau}} \equiv \langle e^{-\tau} \rangle$ (the usual optical depth is related to the absorption coefficient via $d\tau_{\nu}=\alpha_{\nu} dl$). 
For Poisson-distributed absorbers, each of column density $N_{\rm HI}$,
\begin{equation}
\label{taueff poisson expression}
\bar{\tau}(\nu_{0}, z_{0}, z) =
\int_{z_{0}}^{z} dz'
\int_{0}^{\infty}
dN_{\rm HI}
\frac{\partial^{2}N}{\partial N_{\rm HI} \partial z'}
(1 - e^{-\tau_{\nu}}),
\end{equation}
where $\partial^{2}N/\partial N_{\rm HI} \partial z'$ is the column density distribution of intergalactic absorbers versus redshift \citep{1980ApJ...240..387P}. 

The optical depth $\tau_{\nu}$ shortward of the Lyman limit is dominated by the photoelectric opacity of hydrogen and helium,
\begin{equation}
\tau_{\nu} = N_{\rm HI}\sigma_{\rm HI}(\nu) + N_{\rm HeI}\sigma_{\rm HeI}(\nu) + N_{\rm HeII}\sigma_{\rm HeII}(\nu),
\end{equation}
where the $N_{i}$ and $\sigma_{i}$ are the column densities and photoionization cross sections of ion $i$.
While the HI column density distribution is reasonably well determined over large redshift and column density intervals, the incidence of HeI and HeII absorbers is not accurately measured. 
As in previous work, we therefore use a model for the HeI and HeII column densities given the HI column. 

We parameterize the HI column density distribution using broken power laws in $N_{\rm HI}$ and $z$: 
\begin{equation}
\label{eq:column_density_distribution}
f(N_{\rm HI}) \equiv \frac{\partial^{2}N}{\partial z \partial N_{\rm HI}}= A(N_{\rm HI},z) N_{\rm HI}^{-\beta(N_{\rm HI},z)} (1+z)^{\gamma(N_{\rm HI},z)},
\end{equation}
where the power-law parameters $A$, $\beta$, and $\gamma$ are functions of the HI column and redshift, and are constrained so that the distribution is everywhere continuous with respect to both $N_{\rm HI}$ and $z$. 
We adopt a modified version of the broken power-law parameters in Table 1 of \cite{2019MNRAS.485...47P}. 
The \cite{2019MNRAS.485...47P} column density distribution is itself is a modified version of the broken power laws used by HM12. 
This model consists of five column density regimes ($\log{(N_{\rm HI}/{\rm cm^{-2}})}=11-16,~16-18,~18-19.5,~19.5-20.3,$ and $20.3-21.55$) and two redshift regimes ($z\leq 1.56$ and $z> 1.56$).

We use the same parameters as in \cite{2019MNRAS.485...47P} except for the redshift scaling at $z<1.56$. 
While $f(N_{\rm HI}) \propto (1+z)^{0.16}$ at $z\leq1.56$ in HM12 and \cite{2019MNRAS.485...47P}, we assume a stronger redshift scaling $f(N_{\rm HI}) \propto (1+z)$ in this regime (i.e., $\gamma(z \leq 1.56)=1$).  
The motivation for the stronger redshift is that it is more consistent with the observed redshift evolution of Lyman limit systems \citep[][]{2011ApJ...736...42R}, which most directly determine the mean free path of ionizing photons. 
To maintain continuity at $z=1.56$, we renormalize the low-redshift $A$ coefficients from \cite{2019MNRAS.485...47P} by a factor $(1+1.56)^{0.16}/(1+1.56)=0.454$. 
Appendix \ref{sec:fesc_mfp_deg} shows that the HI ionizing mean free path implied by this column density distribution is in excellent agreement with observational constraints.

To obtain HeI and HeII columns from $N_{\rm HI}$, we define the ratios $\eta = N_{\rm HeII}/N_{\rm HI}$ and $\zeta = N_{\rm HeI}/N_{\rm HI}$, which are functions of $N_{\rm HI}$ and the photoionization rates. 
FG09 used radiative transfer calculations to compute $\eta$ and $\zeta$ over a representative range of parameters and improved analytic approximations from \cite{1998AJ....115.2206F}. 
HM12 further improved on these results by using similar radiative transfer calculations but deriving fitting functions for  $\eta$ and $\zeta$ that are more accurate in the limit of optically thick absorbers. 
In this work, we use the improved fitting functions from HM12. 
We have verified, however, that we obtain nearly identical radiative transfer results using the $\eta$ and $\zeta$ fitting functions from FG09 instead. 
This is because, in the limit of optically thick absorbers, the precise values of $N_{\rm HeI}$ and $N_{\rm HeII}$ do not significantly affect the radiative transfer.

For $i \in \{{\rm HI,~HeI,~HeII}\}$, the photoionization rates are defined as
\begin{equation}
\Gamma_{i} = 
4\pi
\int_{\nu_{i}}^{\infty}
\frac{d\nu}{h \nu}
J_{\nu} \sigma_{i}(\nu),
\end{equation}
where $\sigma_{i}$ is the photoionization cross section for the species of interest and $\nu_{i}$ is the frequency at the photoionization edge. 
The photoheating rates are defined similarly:
\begin{equation}
\dot{q}_{i} = 
4\pi
\int_{\nu_{i}}^{\infty}
\frac{d\nu}{h \nu}
J_{\nu} \sigma_{i}(\nu) (h\nu - h\nu_{i}).
\end{equation}

Finally, we note that we have rewritten our radiative transfer solver from F09 to improve both speed and accuracy. 
In F09, we solved the radiative transfer equation on a fixed grid of redshifts and frequencies. 
Because photons redshift as the universe expands, this approach required a very large number of grid points in order to resolve fine structure in rest-frame frequency (e.g., the HeII Ly-series sawtooth). 
We now instead integrate the solution to the radiative transfer equation (\ref{transfer equation solution}) along light cones using a recursive method.
This method allows us to accurately compute $J_{\nu_{0}}$ at any frequency without requiring a large grid that resolves the full spectrum including all the relevant fine-scale structure at higher redshifts. 

\begin{figure}
\begin{center}
\includegraphics[width=0.475\textwidth]{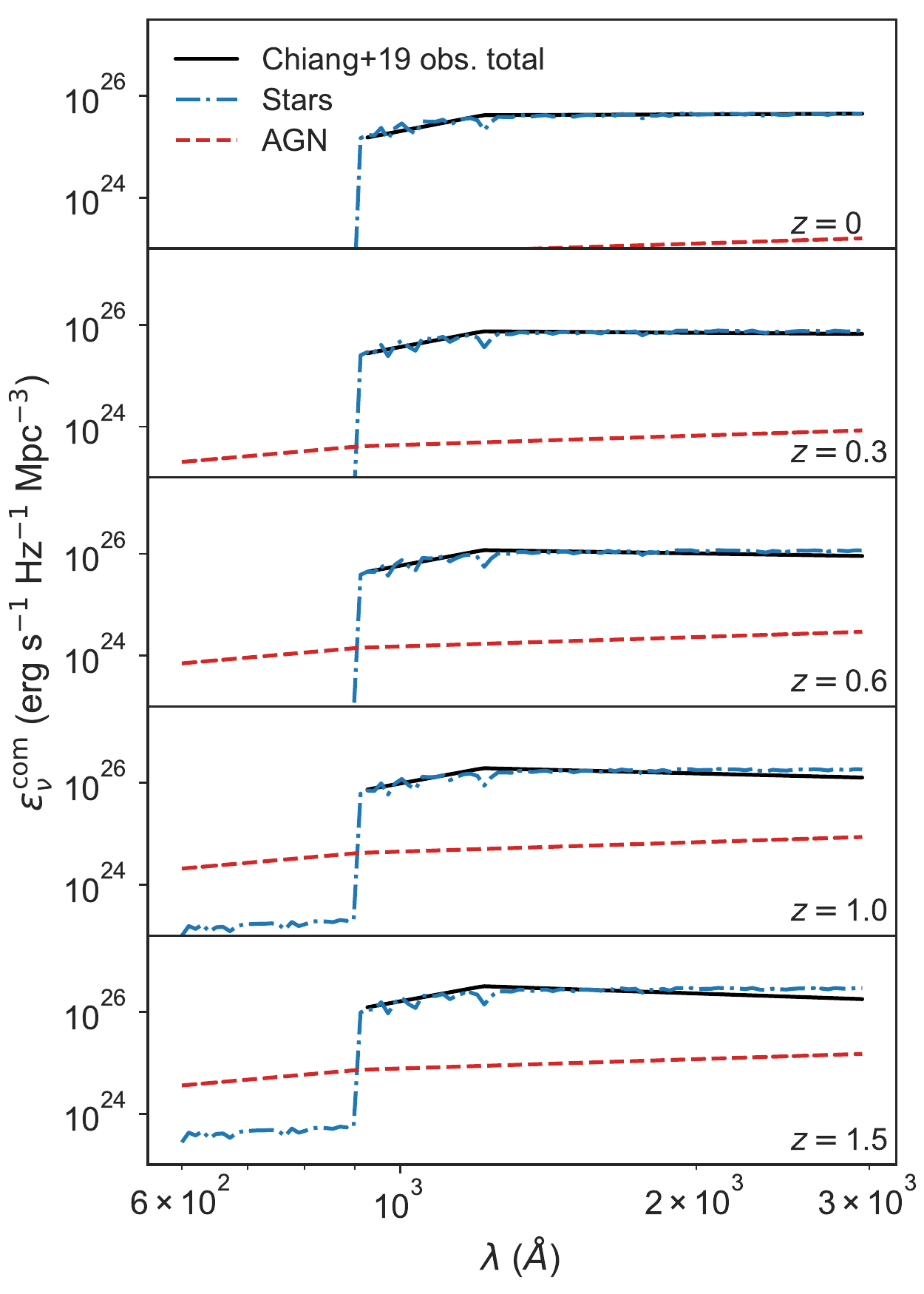}
\end{center}
\caption[]{Frequency-resolved comoving emissivities at low redshift. 
The model stellar (dot-dashes) and AGN (dashes) contributions are compared to an observational inference of the total continuum UV emissivity longward of 912~\AA~obtained by cross-correlating GALEX imaging data with spectroscopic objects from the SDSS \citep[solid;][]{2019ApJ...877..150C}.
The UV emissivity longward of 912~\AA~is dominated by stars at all redshifts, even though the ionizing emissivity is dominated by AGN at all redshifts shown here. 
As in Figure \ref{fig:eps_vs_z}, this is due primarily to different escape fractions assumed for galaxies and AGN, though the intrinsic HI Lyman break in stellar spectra also contributes.
}
\label{fig:epsnu_lowz} 
\end{figure}

\begin{figure*}
\begin{center}
\includegraphics[width=0.98\textwidth]{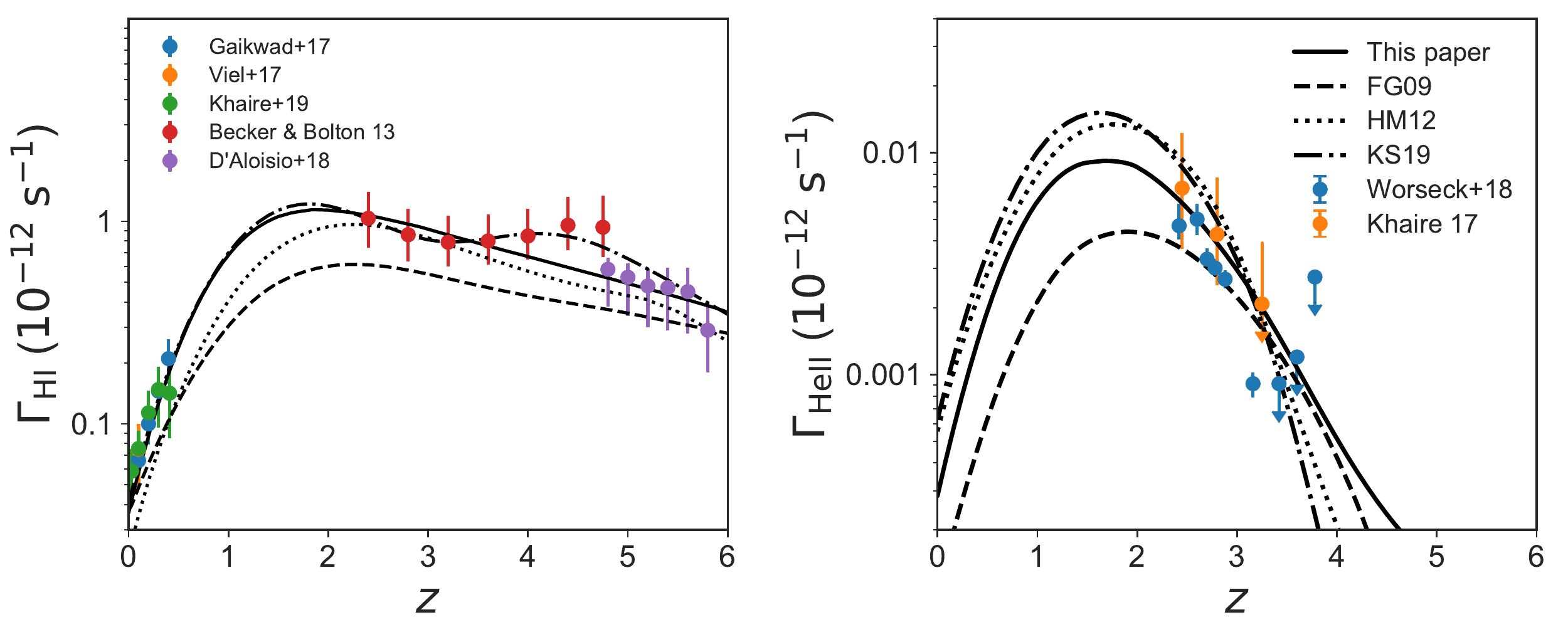}
\end{center}
\caption[]{HI (left) and HeII (right) photoionization rates in the IGM as a function of redshift. 
The solid curves show results from the fiducial UVB synthesis model developed in this work compared to the \citet[][FG09; dashes]{2009ApJ...703.1416F}, \citet[][HM12; dots]{2012ApJ...746..125H}, and fiducial (Q18) \citet[][KS19; dash-dots]{2019MNRAS.484.4174K} models. 
The data points with error bars compile different empirical measurements of the photoionization rates (see \S \ref{sec:integral_constraints}).
}
\label{fig:Gammas_full} 
\end{figure*}

\subsection{Star-forming galaxies}
\label{sec:stellar_spec}
For the evolution of the emissivity from star-forming galaxies, we assume that the rest-UV comoving emissivity at rest wavelength $\approx 1500$~\AA~is given by
\begin{align}
\label{eq:eps_star_1500}
\log_{10}{\epsilon_{\nu1500}^{\star,\rm com} } = (25.62, 25.79, 25.92, 26.04, 26.15, 26.26, \\ \notag
26.32, 26.42, 26.48, 26.54, 26.61, 26.60,\\ \notag
26.60, 26.52, 26.30, 26.11, 25.90, 25.67)
\end{align}
at $z=(0, 0.2, 0.4, 0.6, 0.8, 1, 1.2, 1.4, 1.6, 1.8, 2, 2.1, 3, 3.8,$ $ 4.9, 5.9, 6.8, 7.9)$, and interpolate linearly between these redshifts (emissivity in erg s$^{-1}$ Hz$^{-1}$ Mpc$^{-3}$). 
At $z\leq 2$, these values correspond to the total UV emissivity as function of redshift inferred by \cite{2019ApJ...877..150C} by cross-correlating GALEX UV imaging with sources with spectroscopic redshifts in SDSS. 
The total emissivity at these redshifts is consistent with the emissivity predicted by integrating the galaxy luminosity function down to very faint magnitudes \citep[e.g.,][]{2016ApJ...827..108D} and our results below also indicate that AGN contribute only a small fraction of the total 1500~\AA~emissivity at these redshifts. 
At $z\geq2.1$, the above luminosity densities are inferred from measurements of the rest-UV galaxy luminosity function integrated down to -13.0 AB mag (Bouwens et al., in prep.). 
Relative to the values reported by Bouwens et al., we have adjusted the luminosity density at some redshifts (within uncertainties) to avoid non-monotonic behavior with redshift. 

Figure \ref{fig:eps_vs_z} summarizes the comoving emissivities from star-forming galaxies and AGN adopted in our UVB synthesis model as a function of redshift, at rest-frame wavelengths of 1500~\AA~and 912~\AA. 
We tie our ionizing background models to UV luminosity densities inferred before any correction for dust obscuration because ionizing photons are also absorbed by dust. 
Thus, the ionizing emissivity relevant for modeling the intergalactic background should track the emissivity of UV photons that reach the IGM. 
As \cite{2015ApJ...803...34B} show, the dust-corrected and dust-uncorrected UV luminosity densities diverge increasingly from high to low redshift. 

Galaxies contribute most importantly to the HI photoionization rate at high redshifts $z\gtrsim2$, past which the quasar luminosity function begins to drop much more steeply than the nearly flat photoionization rate \citep[e.g.,][]{2005MNRAS.357.1178B, 2008ApJ...682L...9F, 2008ApJ...688...85F, 2009ApJ...694..842M}. 
We therefore base our fiducial stellar spectral template on recent results from spectroscopic surveys of $z \sim 3$ Lyman break galaxies (LBGs). 
While star-forming populations are observed out to much higher redshifts \citep[e.g.,][]{2017ApJ...843..129B, 2017ApJ...835..113L}, the stellar populations of epoch-of-reionization galaxies have not yet been studied in as much detail as those at $z\sim3$. 
Moreover, absorption by the intergalactic medium makes the direct detection of escaping ionizing radiation from galaxies nearly impossible past $z\sim 3.5$. 

For the spectra of star-forming galaxies, we use templates produced by the BPASS stellar population synthesis model, which includes binary stars \citep[v2.2.1;][]{2017PASA...34...58E}. 
The distributions of binary parameters used in BPASS are as in table 13 of \cite{2017ApJS..230...15M}. 
Our treatment of stellar binaries improves on previous UVB models, which used either simple power-law approximations for spectra (e.g., FG09) or stellar population synthesis models including single stars only (e.g., HM12, P19, KS19). 
Instead of a detailed convolution of the spectral model with a distribution of stellar ages and metallicities, we use fixed templates representative of high-redshift star-forming galaxies, where stars are most important for the UVB. 
Detailed observations of $z\sim2-3.5$ LBGs have recently shown that their nebular emission lines are better modeled by stellar populations including binaries \citep[][]{2014ApJ...796..136B, 2016ApJ...826..159S, 2017ApJ...836..164S}, whose effective temperatures are higher \citep[for an alternative interpretation of observed line ratios based on chemical abundance ratios see, e.g.,][]{2014ApJ...785..153M, 2015ApJ...801...88S}. 
When binaries are included, massive hot stars can persist significantly past the few Myrs predicted by single-star models. 
Support for the importance of stellar binaries at high redshift has also recently emerged from galaxy formation simulations, which find that longer-lived hot stars help boost the effective escape fraction to values high enough to explain HI reionization \citep[e.g.,][]{2015MNRAS.453..960M, 2016MNRAS.459.3614M, 2018MNRAS.479..994R}. 
We use the default `imf135\_300' initial mass fraction (IMF) in BPASS, corresponding to a power-law slope $\alpha_{1}=1.30$ between 0.1~$M_{\odot}$ and 0.5~$M_{\odot}$ and a slope $\alpha_{2}=2.45$ between 0.5~$M_{\odot}$ and 300~$M_{\odot}$ ($dN(<M)/dM \propto M^{\alpha_{i}}$). 
\cite{2018ApJ...869..123S} showed that this choice of IMF yields good fits to LBG spectra. 
We show in Appendix \ref{sec:stellar_fesc} that our results would not change significantly if we instead assumed a \cite{2003ApJ...586L.133C} IMF. 
\begin{figure*}
\begin{center}
\includegraphics[width=0.98\textwidth]{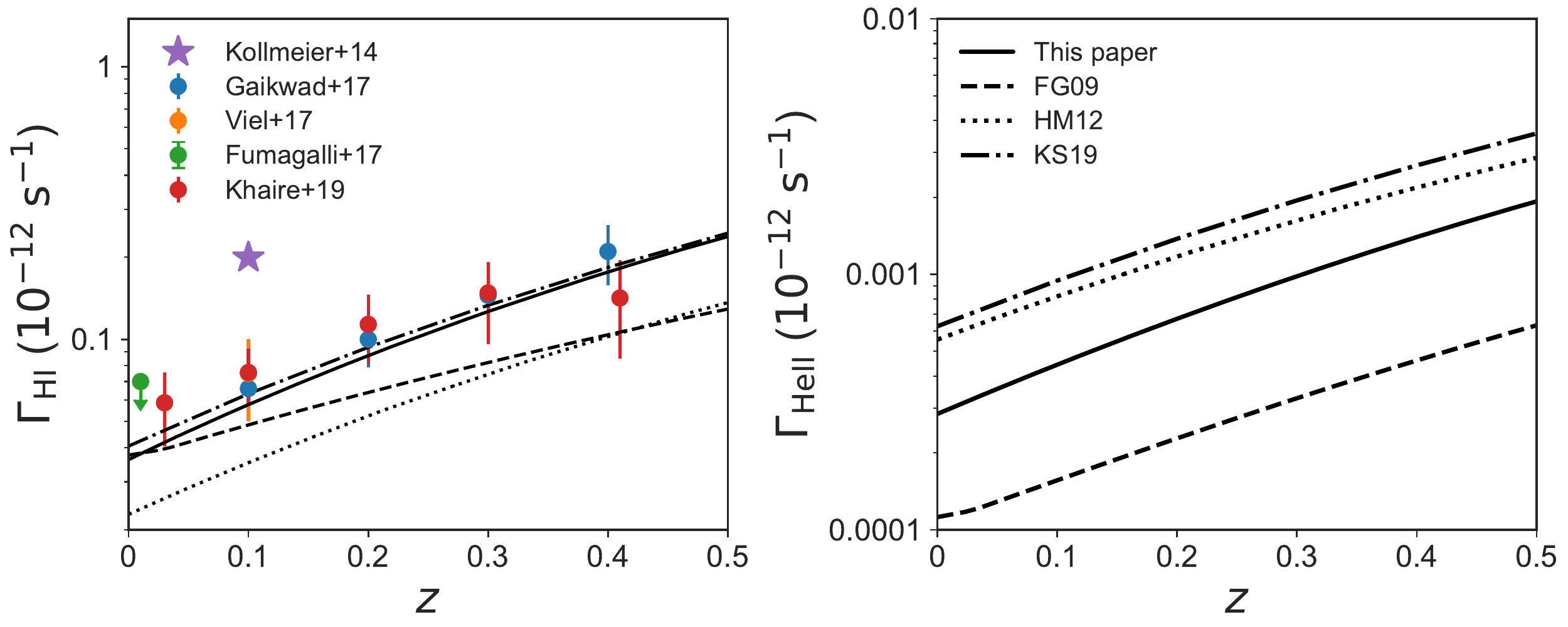}
\end{center}
\caption[]{Same as in Fig. \ref{fig:Gammas_full}, but with redshift range restricted to $z<0.5$ to highlight a regime where our new model differs significantly from HM12 and relevant to the ``photon underproduction crisis'' \citep[purple star measurement;][]{2014ApJ...789L..32K}. The local UVB measurement from \cite{2017MNRAS.467.4802F} is plotted slightly offset from $z=0$ for clarity, and is shown as an upper limit to account for the possibility that the H$\alpha$ recombination emission on which this measurement is based is partially powered by sources other than the UVB. 
}
\label{fig:Gammas_lowz} 
\end{figure*}

Another recent development regarding high-redshift galaxies is the realization that their stellar and gas-phase metallicities can differ substantially.  
\cite{2016ApJ...826..159S} showed that the stellar and nebular spectra of $z\sim2.4$ LBGs can be reconciled if the massive stars have a metallicity $Z_{\star} \approx 0.1$ Z$_{\odot}$  but the nebular gas has a higher metallicity, $Z_{\rm neb} \approx 0.5$ Z$_{\odot}$. 
This apparent discrepancy can be explained by a $\sim 5\times$ supersolar O/Fe abundance for the nebular gas, which is expected for enrichment dominated by core-collapse SNe. 
While O dominates the physics of the nebular gas observed in emission lines, Fe dominates the extreme and far-UV opacity and controls the mass loss rate from stars. 
Thus, the Fe abundance most strongly affects the properties of the massive stars which produce most of the ionizing photons.
For our fiducial stellar template, we therefore assume a low stellar metallicity $Z_{\star} = 0.1$ Z$_{\odot}$ (corresponding to a metal mass fraction $Z=0.002$). 

For dust attenuation, we use the \citet[][]{2016ApJ...828..107R} attenuation curve calibrated to $z\sim3$ observations:
 \begin{equation}
 S_{\nu}^{\rm dust} = S_{\nu}^{\rm intr} 10^{-0.4E(B-V)k(\lambda)},
 \end{equation}
 where $k(\lambda)$ is given in \cite{2015ApJ...806..259R} for $0.15 \leq \lambda \leq 2.85$ $\mu$m and in \cite{2016ApJ...828..107R} for $\lambda <0.15$ $\mu$m. 
 The \citet[][]{2016ApJ...828..107R} attenuation curve provides a better match to observations of high-redshift galaxies than the \cite{2000ApJ...533..682C} attenuation curve, and implies a lower attenuation in the far-UV for a given $E(B-V)$. 
$S_{\nu}^{\rm intr}$ is the intrinsic spectrum of the stellar population before any dust correction is applied and $S_{\nu}^{\rm dust}$ is the same spectrum but attenuated by dust. 

For the escape fraction of ionizing photons ($f_{\rm esc}^{\star}$), we use the ``holes'' model in which $f_{\rm esc}^{\star}$ is determined by the fraction of sight lines from hot stars along which photons can escape unimpeded, while ionizing photons are completely absorbed along other sight lines. 
We furthermore assume no dust obscuration along holes. 
A holes model is supported by high-resolution galaxy formation simulations \citep[e.g.,][]{2013ApJ...775..109K, 2015ApJ...801L..25C, 2015MNRAS.453..960M} as well as observations \citep[e.g.,][]{2018ApJ...869..123S, 2019ApJ...878...87F}. 
The net angle-averaged emergent spectrum from galaxies is thus
\begin{equation}
\label{eq:holes_model}
S_{\nu} =
\left\{
\begin{array}{ll}
S_{\nu}^{\rm dust}(1-f_{\rm esc}^{\star})  +  S_{\nu}^{\rm intr} f_{\rm esc}^{\star} & \nu < {\rm1~Ry} \\
S_{\nu}^{\rm intr} f_{\rm esc}^{\star} & \nu \geq {\rm1~Ry} 
\end{array}
\right. .
\end{equation}
Figure \ref{fig:BPASS} shows our stellar spectral template for fiducial $E(B-V)$ and $f_{\rm esc}^{\star}$ values. 
We fix $E(B-V)=0.129$ in our calculations, a value which \cite{2018ApJ...869..123S} found to provide a good fit to a composite $z\sim3$ LBG spectrum. 
The stellar spectra assume a continuous star formation history with an age of 300 Myr, which is appropriate for LBGs \citep[e.g.,][]{2012ApJ...754...25R}. 
The exact age of the stellar population is not critical as the stellar UV flux becomes approximately constant after $\sim 100$ Myr of continuous star formation. 
The normalization of the spectral template in equation (\ref{eq:holes_model}) is arbitrary; at any given redshift, the spectral \emph{shape} is used to evaluate the emissivity at any frequency $\nu$ given the prescribed 1500~\AA~emissivity (eq \ref{eq:eps_star_1500}).
\begin{figure*}
\begin{center}
\mbox{
\includegraphics[width=0.48\textwidth]{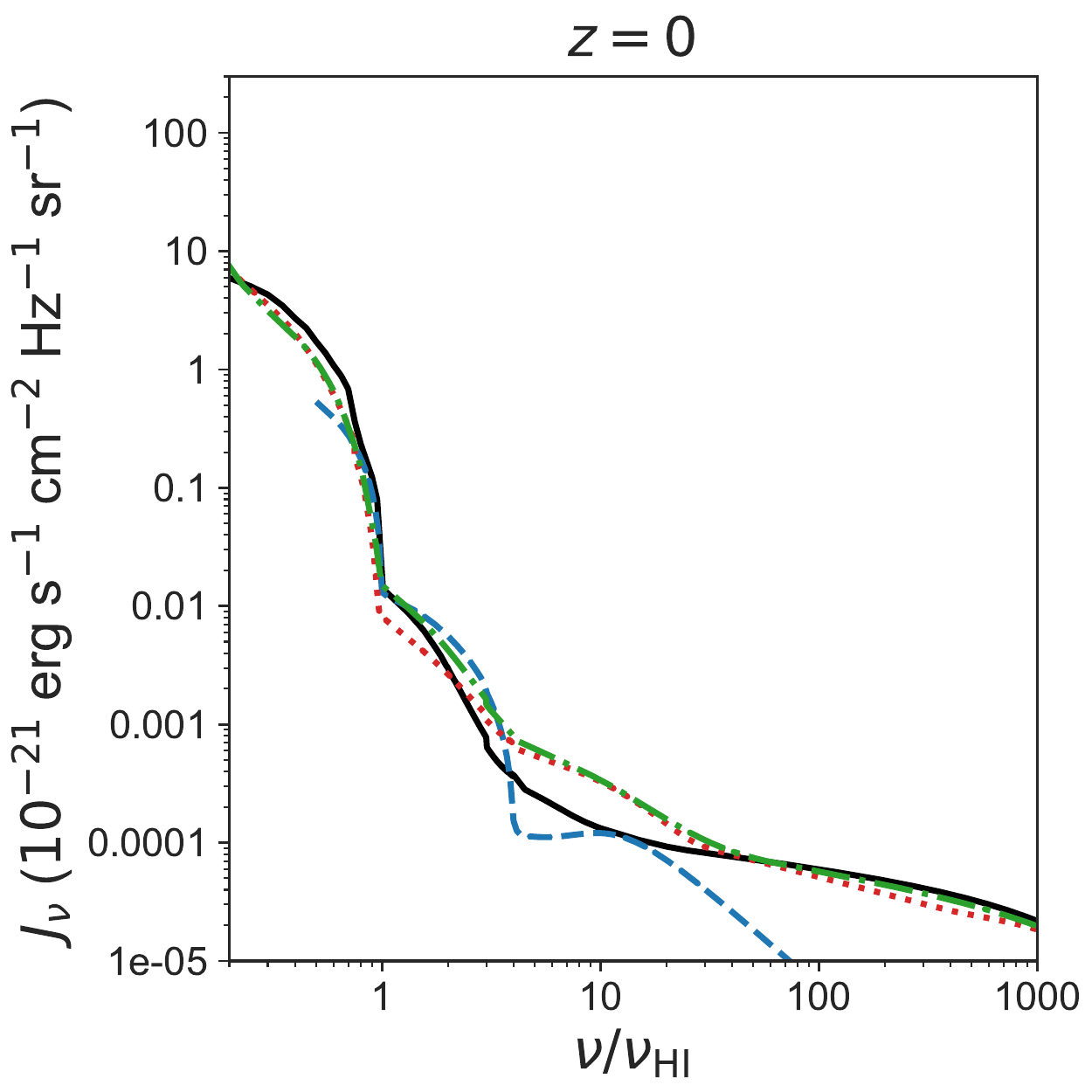}
\includegraphics[width=0.48\textwidth]{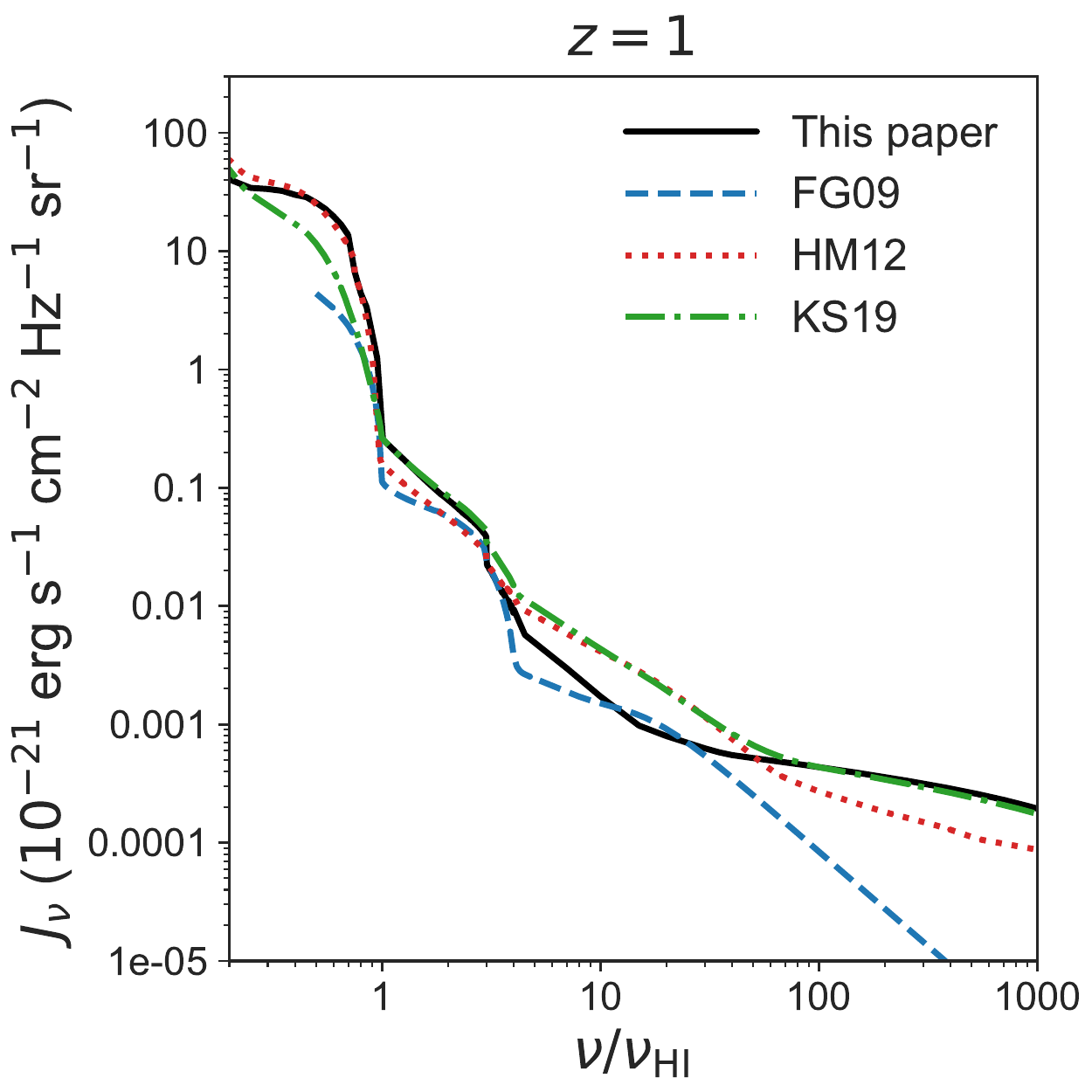}
}
\mbox{
\includegraphics[width=0.48\textwidth]{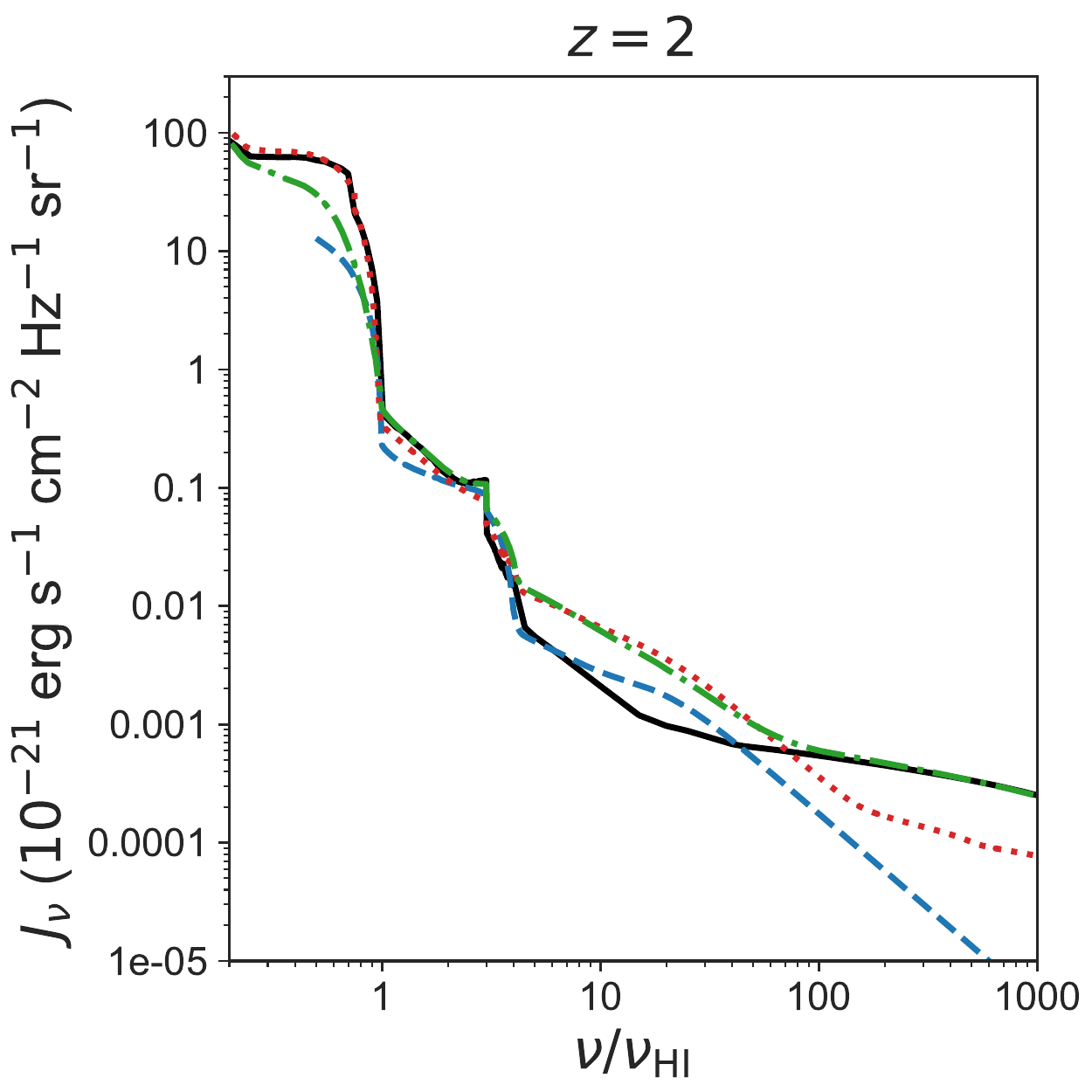}
\includegraphics[width=0.48\textwidth]{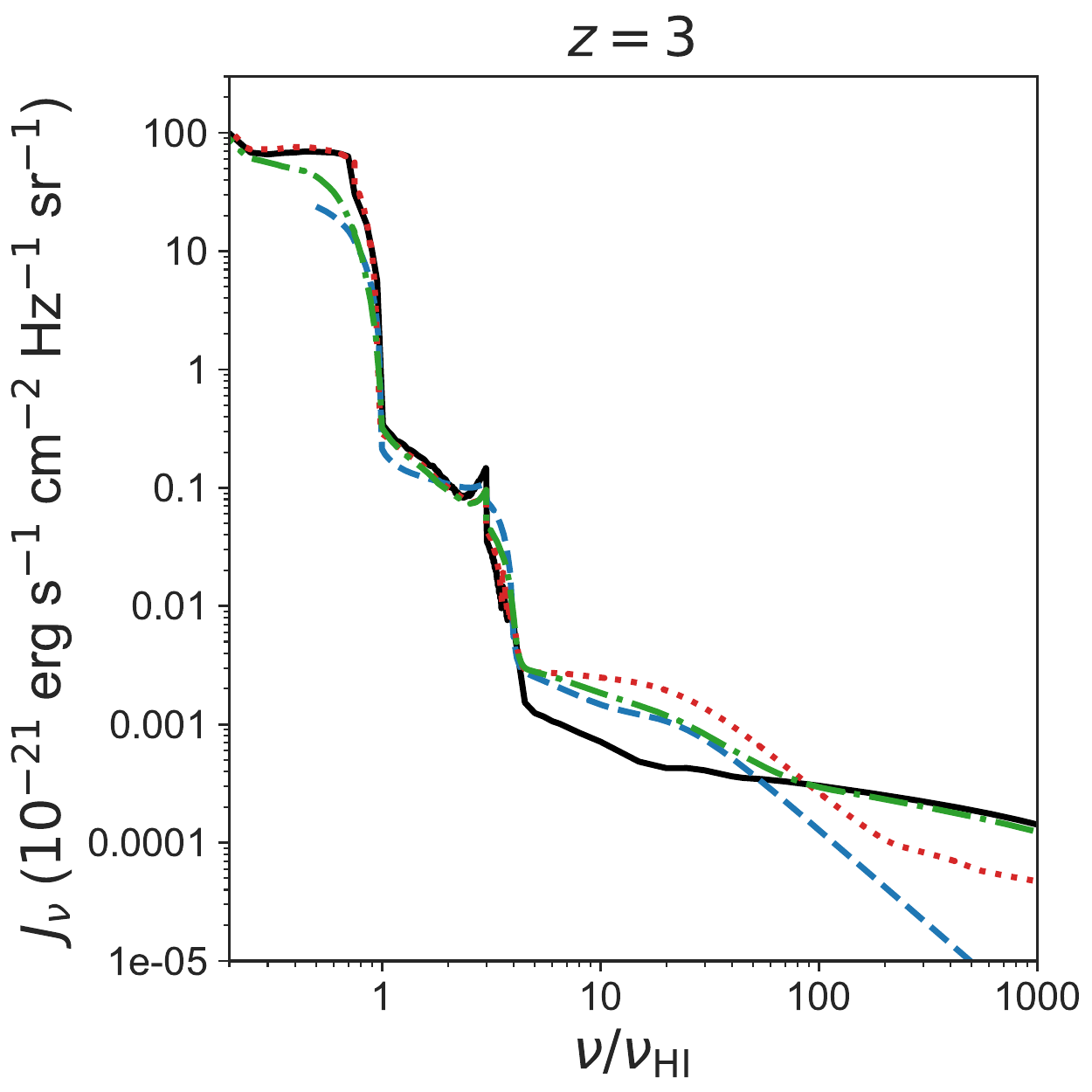}
}
\end{center}
\caption[]{Spectrally-resolved cosmic UV/X-ray background models at different redshifts. 
We show the isotropic specific intensity $J_{\nu}$ as a function of frequency in units of the HI ionization potential (1 Ry=13.6 eV), so the maximum frequency shown corresponds to 13.6 keV). 
The solid curves show the results from this paper. 
These are compared to the previous FG09, HM12, and KS19 UVB synthesis models.
}
\label{fig:Jnu} 
\end{figure*}

The escape fraction is a major uncertainty in models of the ionizing background. In detail, it likely depends on galaxy properties, including redshift and mass \citep[e.g.,][]{2008ApJ...672..765G, 2016ApJ...833...84X}. 
For our purpose, we require an ``effective'' escape fraction weighted by the intrinsic emissivity of the sources. 

By combining constraints on the ionization of the IGM at $z\sim3$ from the Ly$\alpha$ forest and the observed UV galaxy luminosity function, \cite{2012MNRAS.423..862K} found that reconciling the galaxy UV luminosity function with the reionization history inferred from the WMAP-7 optical depth could be achieved by assuming an escape fraction that increases with increasing redshift (other options are to invoke a large population of galaxies below detection limits, or a substantially higher ionizing efficiency at very high redshift). 
A similar conclusion is supported by more recent data. 
For example, \cite{2015ApJ...802L..19R} assume a fiducial escape fraction of 20\% in $z\gtrsim 6$ reionization models constrained by HST and Planck observations. 
Furthermore, the most recent UVB synthesis models all assume a galaxy escape fraction that increases with redshift, albeit with different functional forms \citep[][]{2012ApJ...746..125H, 2019MNRAS.485...47P, 2019MNRAS.484.4174K}.

Motivated by these results, we assume 
an effective escape fraction that increases continuously with redshift but constrained to stay 
under 20\%:\footnote{This expression implies that the modeled galaxy escape fraction can be very low at $z\sim0$.
Although the constraints on the escape fraction remain relatively poor in this regime (direct Lyman continuum detection requires space-based observations), this is consistent with the fact that many previous attempts to detect escaping ionizing radiation at low redshift have primarily yielded upper limits \citep[e.g.,][]{1995ApJ...454L..19L, 2001ApJ...558...56H}. 
Significant escaping ionizing radiation has been detected from some 
low-redshift galaxies \citep[see the compilation of HST/COS detections in][]{2018A&A...616A..30C}, but the existing detections tend to be biased toward extreme starburst galaxies with properties similar to LBGs. 
\cite{2011ApJ...730....5H} proposed that extreme feedback by a powerful starburst may be required to enable the escape of ionizing radiation from galaxies. 
High star formation rates and powerful galactic winds are common at high redshift but much rarer in the nearby Universe, which could explain strong redshift evolution in the escape fraction.
}
\begin{equation}
\label{eq:fesc_stars}
f_{\rm esc}^{\star} = \min \left( f_{\rm esc}^{\star,z=3} \left( \frac{1+z}{4} \right)^{2.5},~0.2 \right).
\end{equation} 
For our assumed emissivity and IGM properties, we find that $f_{\rm esc}^{\star,z=3}=0.01$ provides a good match to the HI photoionization rate inferred from the Ly$\alpha$ forest at $z \gtrsim 3$ (see \S \ref{sec:integral_constraints}). 
This value is similar to the absolute escape fraction inferred by other studies by comparing the $\approx 1500$~\AA~(non-ionizing) galaxy emissivity to the total HI photoionization rate at $z\gtrsim3$. 
In particular, after accounting for differences in definitions, \cite{2008ApJ...688...85F}, \cite{2012ApJ...746..125H}, \cite{2019MNRAS.485...47P}, and \cite{2019MNRAS.484.4174K} all find that the integrated observational constraints can be matched assuming $f_{\rm esc}^{\star,z=3} \approx 0.01-0.02$.

It is interesting, however, that the absolute escape fraction inferred from integral constraints at $z=3$ is substantially lower than indicated by recent ``direct'' measurements of the escape fraction from $z\approx3$ star-forming galaxies. 
\cite{2018ApJ...869..123S} carried out a comprehensive analysis of a sample of $z\approx3$ LBGs and found an average absolute escape fraction of $9\pm1$\%. 
In an HST survey $z \approx 3.1$ galaxies, \cite{2019ApJ...878...87F} found an average escape fraction (accounting for non-detections) of $\approx 5$\%. 
Appendix \ref{sec:stellar_fesc} discusses some effects that may contribute to the difference, but the resolution is not clear. 
One possibility is that the higher escape fractions inferred from direct observations 
are not representative of the UV emissivity-weighted escape fraction. 
This is plausible since, for example, the state-of-the-art \cite{2018ApJ...869..123S} and \cite{2019ApJ...878...87F} studies only detect escaping ionizing photons from relatively luminous galaxies with $L_{\rm UV}\gtrsim0.25 L^{\star}_{\rm UV}$.  
For a \cite{1976ApJ...203..297S} luminosity function with a faint-end slope $\alpha \approx -1.7$, such galaxies account for only $\approx 30\%$ of the total UV emissivity.

At face value, this ``incompleteness'' effect goes in the direction \emph{opposite} to what is needed to explain the discrepancy, since observations indicate that $f_{\rm esc}^{\star}$ correlates positively with the Ly$\alpha$ equivalent width $W({\rm Ly\alpha})$, while $W({\rm Ly\alpha})$ tends to increase with \emph{decreasing} $L_{\rm UV}$ \citep[][]{2018ApJ...869..123S}. 
This suggests that, if anything, on average fainter galaxies should have higher escape fractions. 
Nonetheless, accounting for this $W({\rm Ly\alpha})$ dependence, \cite{2018ApJ...869..123S} estimate that $L_{\rm UV}\gtrsim0.25 L^{\star}_{\rm UV}$ galaxies \emph{alone}  produce a comoving ionizing emissivity $\epsilon_{\nu 912}^{\rm \star,com} \approx 5.9 \times 10^{24}$ erg s$^{-1}$ Hz$^{-1}$ Mpc$^{-3}$ at $z \approx 3$, which is almost exactly equal to the \emph{total} ionizing emissivity needed to match the Ly$\alpha$ forest-inferred $\Gamma_{\rm HI}$ at $z=3$ in our UVB synthesis model (see Fig. \ref{fig:eps_vs_z}). 
There is thus, at face value, no room for $L_{\rm UV}\lesssim0.25 L^{\star}_{\rm UV}$ galaxies to produce a significant amount of escaping ionizing photons. 
If the ionizing emissivity predicted using the \cite{2020arXiv200102696S} AGN luminosity function in the next section is accurate at $z=3$, AGN contribute about half of the total ionizing emissivity at this redshift, and a galaxy escape fraction $\approx 9\%$ may overproduce $\Gamma_{\rm HI}(z=3)$ even if restricted to $L_{\rm UV}\gtrsim 0.25 L^{\star}_{\rm UV}$ galaxies. 
We note, however, that \cite{2020MNRAS.tmpL...7B} did not detect any escaping ionizing photons in a recent study of faint ($L_{\rm UV}\approx 0.1 L^{\star}_{\rm UV}$), $z\approx3.1$ Ly$\alpha$ emitters. 
\cite{2020MNRAS.tmpL...7B} conclude that this result is in tension with an extrapolation of \cite{2018ApJ...869..123S}'s relation between $f_{\rm esc}^{\star}$ and $W({\rm Ly\alpha})$ to their sample of high-$W({\rm Ly\alpha})$ galaxies. 
Thus, it is possible that faint galaxies have lower than escape fractions than suggested by trends measured in samples of brighter galaxies.

\subsection{Active galactic nuclei}
\label{sec:agn_spec}
In FG09, we used the bolometric luminosity function model from  \citet[][hereafter HRH07]{2007ApJ...654..731H}, which was based on a large compilation of obervational data available at the time, to evaluate the rest-frame AGN emissivity at 4400~\AA~($B$ band). 
We then adopted simple power-law approximations for the AGN spectrum from the optical to the extreme UV. 
In this paper, we update our AGN treatment to account for new luminosity function measurements (especially at high redshift, where the HRH07 luminosity function relied heavily on extrapolations) and with an improved spectral model including both unobscured and obscured sources. 

We use the new analysis of the AGN bolometric luminosity function from \cite{2020arXiv200102696S}. This new analysis is an update of the HRH07 bolometric luminosity function which includes a large number of new AGN measurements that have become available since HRH07's study. 
We start with the AGN ionizing emissivity derived by Shen et al. by integrating their luminosity function  (at the rest wavelength 912~\AA). 
For the comoving emissivity in erg s$^{-1}$ Hz$^{-1}$ Mpc$^{-3}$, this gives:
\begin{align}
\label{eq:shen_eps_fit}
\epsilon_{\nu 912}^{\rm AGN,com} = 10^{24.11} (1+z)^{5.87} \frac{\exp{(0.73z)}}{\exp{(3.06z)} + 15.60}.
\end{align}
A spectral template for the AGN population is then used to evaluate the emissivity at other wavelengths.

We derive and use an AGN spectral template that accounts for both unobscured and obscured sources. 
For unobscured (intrinsic) AGN spectra, our template closely follows that used by \cite{2020arXiv200102696S} in their analysis of the bolometric luminosity function. Briefly:
\begin{itemize} 
\item {\bf Optical-UV ($\lambda \geq 912$~\AA):} The \cite{2013ApJS..206....4K} template for the mean intrinsic quasar spectrum based on $\sim100,000$ spectra at $0.064 < z < 5.46$ (the template extends into the infrared, but we do not use that part in this work). 
\item {\bf Ionizing UV ($600<\lambda \leq 912$~\AA):} A power-law $F_{\nu} \propto \nu^{\alpha}$, where $\alpha=-1.7$ based on a stack of $z\sim 2.4$ quasars corrected for intervening Ly$\alpha$ forest and Lyman continuum absorption \citep{2015MNRAS.449.4204L}.
\item {\bf X-rays ($>0.5$ keV):} An exponentially-truncated power-law template with photon index $\Gamma$ and cut-off energy $E_{\rm c}$: $F_{E}/E \propto E^{-\Gamma} \exp{(-E/E_{\rm c})}$. We set $\Gamma=1.8$ and $E_{\rm c}=300$ keV, which are representative of the best fits from several X-ray studies \citep[e.g.,][]{2008A&A...485..417D, 2014ApJ...786..104U, 2015MNRAS.451.1892A}. 
To this, we add a Compton reflection component using the PEXRAV model \citep[][]{1995MNRAS.273..837M}, assuming a 
reflection strength of unity, an inclination angle of $60^{\circ}$, and solar abundances. 
\item {\bf Extreme UV to X-rays (600~\AA~to 0.5 keV):} We assume a single power law in the energy `gap' between the UV and X-ray components defined above. 
The power law connects the template fluxes at 600~\AA~and 0.5 keV, where the relative amplitude of the X-ray component relative to the ionizing UV component is set to match observational measurements of the 
X-ray-to-optical ratio, $\alpha_{\rm OX}$, given the optical-to-UV and X-ray spectral shapes. 
Let $L_{\nu}(2~{\rm keV})$ and $L_{\nu}(2500~{\rm \AA})$ be specific luminosities in units of erg s$^{-1}$ Hz$^{-1}$. 
Observations have found a correlation of the form $\log{L_{\nu}(2~{\rm keV})} = B \log{L_{\nu}(2500~{\rm \AA})} + C$, which implies a luminosity-dependent $\alpha_{\rm OX} =$ 
$\frac{
\log{L_{\nu}(2~{\rm keV})}-\log{L_{\nu}(2500~{\rm \AA})}
}
{
\log{\nu({\rm 2~keV})}-\log{\nu(2500~{\rm \AA})}
}$
$=0.384 \log{\left( \frac{L_{\nu}(2~{\rm keV})}{L_{\nu}(2500~{\rm \AA})} \right)}$. Since $\alpha_{\rm OX}$ is luminosity-dependent, we must choose a representative AGN luminosity to set the relative amplitude of the UV and X-ray components. 
We use the value of  $L_{\nu}(2500~{\rm \AA})$ corresponding to a $B$-band optical luminosity $L_{\rm B} = \nu L_{\nu} |_{\rm 4400~\AA}=10^{43.3}$ erg s$^{-1}$. 
This is roughly the optical luminosity of AGN that contribute most to the AGN UV emissivity at $z=0.25$ (see \S \ref{sec:agn_spec}). 
\end{itemize}

One approach to obtain a spectral template for obscured AGN is to convolve the intrinsic AGN spectrum with a distribution of obscuring columns, $N_{\rm H}$, making assumptions about the dust content of obscuring material. 
This is, for example, what is done to model obscured AGN in the bolometric luminosity function analyses of HRH07 and \cite{2020arXiv200102696S}. 
We follow a different approach, similar to what was done by \citet[][hereafter SOS04]{2004MNRAS.347..144S} to construct an older AGN spectral template. 
The idea is to assume that the CXB (at $z=0$) above some minimum energy is dominated by a sum over all AGN (obscured and non-obscured), while the optical-UV background is dominated by non-obscured AGN.  
For a given AGN luminosity function, we can then derive what X-ray spectral shape is needed to match the observed CXB. 

More specifically, we define the following SED for the total AGN population averaged over unobscured and obscured sources: 
\begin{align}
F_{E}^{\rm tot} = (1-f_{\rm obsc}) F_{E}^{\rm int} + f_{\rm obsc} F_{E}^{\rm obsc},
\end{align}
where $f_{\rm obsc}$ is the obscured fraction, $F_{E}^{\rm int}$ is the mean intrinsic AGN SED defined above, and $F_{E}^{\rm obsc}$ is a template for the mean SED of obscured AGN. 
For simplicity, we first neglect Compton thick AGN ($N_{\rm H}>10^{24}$ cm$^{-2}$) and normalize the templates such that $F_{E}^{\rm int}({\rm 50 keV})=F_{E}^{\rm obsc}({\rm 50 keV})=F_{E}^{\rm tot}({\rm 50 keV})$ (i.e., we neglect AGN that are significantly absorbed at 50 keV). 
We can then solve for the obscured AGN template:
\begin{align}
F_{\rm E}^{\rm obsc} = \frac{F_{\rm E}^{\rm tot} - (1-f_{\rm obsc}F_{\rm E}^{\rm int})}{f_{\rm obsc}}.
\end{align}

We use the fact that obscured AGN contribute non-negligibly only to X-rays and higher energies, and that both obscured and non-obscured sources contribute to the CXB. 
Furthermore, we assume \citep[as observations indicate, e.g.,][]{2006ApJ...645...95H, 2017ApJ...837...19C} that most of the $\gtrsim 2$ keV CXB is produced by AGN. 
As in SOS04, we thus have the requirement that the total AGN spectral template $F_{E}^{\rm tot}$ should produce the measured CXB after integrating over the AGN luminosity function and redshifting. 
We use a functional form based on SOS04 for the total AGN template in the X-ray regime: 
\begin{equation}
\label{eq:SOS04_HE}
F_{E}^{\rm tot} = A
\left\{
\begin{array}{ll}
E^{-\alpha} e^{-E/E_{1}}& 2~{\rm keV}\leq E < E_{0} \\
B(1+k E^{\beta-\gamma}) E^{-\beta} & E \geq E_{0}\equiv(\beta-\alpha)E_{1}.
\end{array}
\right. 
\end{equation}
SOS04 found that the following parameters provided a good fit to a variety of observational constraints on AGN spectra, including the CXB: 
$A=2^{\alpha} e^{2/E_{1}}$,~$B=E_{0}^{\beta-\alpha} e^{-(\beta-\alpha) / (1+k E_{0}^{\beta-\alpha})}$, $\alpha=0.24$, $\beta=1.6$, $\gamma = 1.06$, $E_{1}=83$ keV, $k=4.1\times 10^{-3}$. 
For our UVB synthesis model, we adopt the same parameters as SOS04, except for $E_{1}$ which we set to 67 keV. 
The different value is due to the fact that we assume a different redshift evolution for the AGN luminosity function in this work, so different spectral parameters are necessary to produce a good fit to the CXB (we compare our model results to the observed CXB in \S \ref{sec:integral_constraints}). 
Figure \ref{fig:AGN_spectra} shows the resulting AGN spectral template.

We find that $f_{\rm obsc}=0.75$ provides an excellent match to the relative intensities of the UV and X-ray backgrounds. 
Since we normalized the obscured template such that it matches the intrinsic template at 50~keV, $f_{\rm obsc}$ may be roughly interpreted as the fraction of AGN that are obscured in the UV (or optical) but not at 50 keV. 
Let us call these the Type II AGN. 
According to this definition, Type II AGN exclude the most heavily obscured, Compton-thick AGN ($N_{\rm H} > 10^{24}$ cm$^{-2}$), which are needed to explain the shape of the CXB \citep[e.g.,][]{2007A&A...463...79G}. 
Although some CXB models imply that Compton-thick AGN are roughly as abundant as Type II AGN \citep[e.g.,][]{2014ApJ...786..104U}, which would perhaps imply an uncomfortably high total obscured fraction, the inferred abundance of Compton-thick AGN from CXB modeling is degenerate with the assumed strength of the X-ray reflection component from the total AGN population. 
Recent studies based on the direct detection and modeling of heavily obscured AGN in hard X-rays using the NuSTAR satellite however indicate a much lower fraction of Compton-thick AGN, corresponding to $\sim10-20\%$ of the Type II population \citep{2018ApJ...867..162M, 2019A&A...621A..28G}.

We also considered the AGN UV luminosity function recently derived by \cite{2019MNRAS.488.1035K} from a compilation of different surveys. 
However, unlike the \cite{2020arXiv200102696S} model, the \cite{2019MNRAS.488.1035K} luminosity function has a faint-end slope so steep that the total emissivity is sensitive to the limiting magnitude of the integral (e.g., $M_{\rm lim}=-18$ vs. $M_{\rm lim}=-21$), i.e. the abundance of very faint AGN for which direct observational constraints are poor. 
We compare the ionizing emissivities implied by the HRH07, \cite{2019MNRAS.488.1035K}, and \cite{2020arXiv200102696S} AGN luminosity functions and expand on the issue of completeness in Appendix \ref{sec:AGN_appendix}. 

\subsection{Recombination emission}
\label{sec:recombs}
For the recombination emissivity, we include Lyman continuum (LyC; 1 Ry) and Ly$\alpha$ (0.75 Ry) recombination emission from HI; and LyC (4 Ry), Balmer continuum (BalC; 1 Ry), and Ly$\alpha$ recombination emission from HeII (3 Ry). 
For HI LyC and HeII LyC, BalC, and Ly$\alpha$, we use the approximations developed in FG09. 

For HI Ly$\alpha$, we assume that 0.68 HI Ly$\alpha$ photon is produced for each ionizing photon absorbed in the ISM of a star-forming galaxy ($\propto(1-f_{\rm esc}^{\star}$)), corresponding to case B recombination. 
We then assume that a fraction of these Ly$\alpha$ photons are destroyed by dust before escaping the galaxy by applying the same dust attenuation that we use to attenuate the UV continuum of galaxies. 
At any redshift, the emissivity from this process is modeled as a spatially-homogeneous $\delta$-function in frequency. 
Note that this is only a crude approximation for HI Ly$\alpha$ as it neglects resonant scattering effects, which can strongly affect the radiative transfer in this line \citep[e.g.,][]{2006ApJ...649...14D, 2006A&A...460..397V, 2010ApJ...725..633F, 2019MNRAS.484...39S}. 
Moreover, we neglect Ly$\alpha$ photons produced by quasars and by recombinations from intergalactic clouds. 

Relative to FG09, we have improved our radiative transfer code to include ``sawtooth'' absorption by HeII Lyman series (Lys) lines between 3 and 4 Ry and the corresponding recombination emission following \cite{2009ApJ...693L.100M}. 
\begin{figure}
\begin{center}
\includegraphics[width=0.47\textwidth]{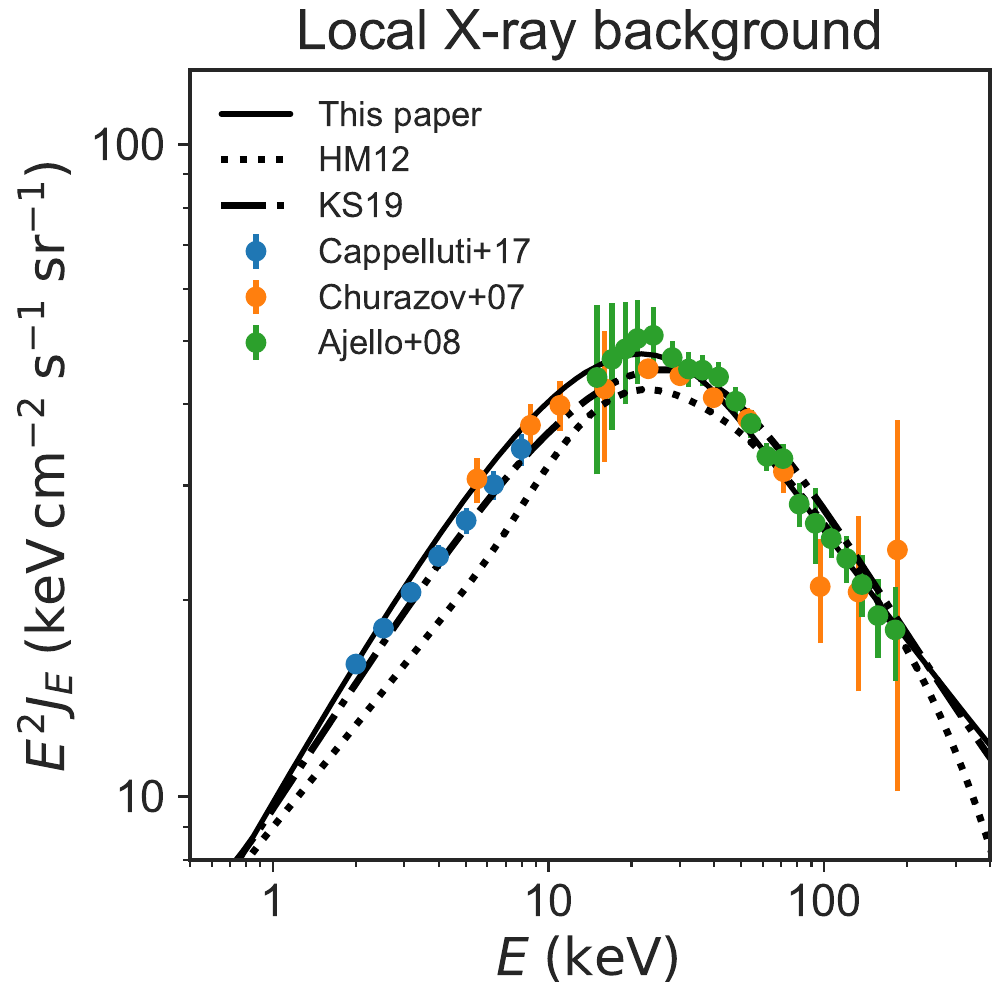}
\end{center}
\caption[]{Representative measurements of the local X-ray background compared to our UV/X-ray background model at $z=0$ (solid), the HM12 model (dots), and the KS19 model (dash-dots). 
The \cite{2017ApJ...837...19C} data points (blue) show the X-ray background at 2-7 keV measured using the Chandra X-ray Observatory, the \cite{2007A&A...467..529C} data points (orange) show an INTEGRAL measurement bracketing the CXB peak at $\sim$20-30 keV, and the \cite{2008ApJ...689..666A} data points (green) show a more precise Swift BAT hard X-ray measurement extending to 200 keV.
}
\label{fig:CXB} 
\end{figure}

\subsection{Integral constraints and synthesis results}
\label{sec:integral_constraints}
The most stringent constraints on our UVB model are integral constraints that do not require detailed assumptions about the radiation sources. 
In this section, we compare our new synthesis model with several integral constraints as well as to the previous FG09, HM12, and fiducial KS19 model (their Q18 model). 
In \S \ref{sec:reionization}, we will also compare our new model to results from \cite{2017ApJ...837..106O} and \cite{2019MNRAS.485...47P} on modeling the effects of reionization.

Figure \ref{fig:epsnu_lowz} compares the frequency-resolved total UV emissivity at $z\leq 1.5$ inferred by \cite{2019ApJ...877..150C} 
to the model emissivities for galaxies and AGN based on the assumptions detailed in \S \ref{sec:stellar_spec}-\ref{sec:agn_spec}. 
Interestingly, the spectral shape of the total emissivity 
is in excellent agreement with the BPASS stellar template used in our UVB model (including an apparent but subtle change in spectral slope around 1216~\AA), providing support for this stellar template. 

Figures \ref{fig:Gammas_full} and \ref{fig:Gammas_lowz} compile $\Gamma_{\rm HI}$ \citep[][]{2013MNRAS.436.1023B, 2014ApJ...789L..32K, 2017MNRAS.466..838G, 2017MNRAS.467L..86V, 2018MNRAS.473..560D, 2019MNRAS.486..769K} and $\Gamma_{\rm HeII}$ \citep[][]{2017MNRAS.471..255K, 2018arXiv180805247W} measurements from quasar absorption spectra. 
These constraints include contributions from all ionizing sources, even galaxies or AGN too faint to be individually detected. 
We also include an estimate of $\Gamma_{\rm HI}$ at $z=0$ based on modeling H$\alpha$ fluorescence emission from a nearby HI disk \citep[][]{2017MNRAS.467.4802F}.

Figure \ref{fig:Jnu} shows the spectrally-resolved model background intensity  
at selected redshifts ($z=0,~1,~2$, and $3$) and compares to the previous FG09, HM12, and KS19 UVB models. 
In Figure \ref{fig:CXB}, we compare the models with measurements of the local X-ray background from \citet[][using Chandra]{2017ApJ...837...19C}, \citet[][using INTEGRAL]{2007A&A...467..529C}, and \citet[][using Swift BAT]{2008ApJ...689..666A}.

Overall, our new UVB model provides an excellent simultaneous match to the empirical constraints considered. 
This is in large part because we calibrated some model parameters to satisfy these constraints, but it is noteworthy that we were able to so by including only radiation from galaxy and AGN populations consistent with observed luminosity functions, and assuming well-motivated spectral templates for each.

One apparent exception is $\Gamma_{\rm HI}$ at $z\sim4.5-5$, which our UVB model appears to underestimate relative to the inference from the Ly$\alpha$ forest by \cite{2013MNRAS.436.1023B}. 
However, our synthesis model agrees well with the measurements from \cite{2018MNRAS.473..560D}, which overlap at $z\sim5$ with the redshift range covered by \cite{2013MNRAS.436.1023B}. 
Since the discrepancy suggests that either the \cite{2013MNRAS.436.1023B} or the \cite{2018MNRAS.473..560D} data points are affected by systematic effects at $z\sim5$, we do not require our model to match the discrepant points from \cite{2013MNRAS.436.1023B}. 

It is also noteworthy that our model provides a good match to most $\Gamma_{\rm HI}$ observational constraints at $z<0.5$. 
In this regime, the UVB in our model is dominated by AGN whose contribution is calculated using relatively standard assumptions: an AGN luminosity function and spectral template calibrated to the latest data (\S \ref{sec:agn_spec}), and an unity escape fraction for AGN. 
Thus, our UVB model does not appear to require non-standard ionizing sources or modeling assumptions to explain the low-redshift HI ionizing background \citep[cf.][]{2014ApJ...789L..32K}.
  
\cite{2014ApJ...789L..32K} reached a different conclusion by comparing their $\Gamma_{\rm HI}$ measurement plotted in Figure \ref{fig:Gammas_lowz} to the HM12 synthesis model. 
Our results suggest that the ``photon underproduction crisis'' noted by \cite{2014ApJ...789L..32K} can be resolved by a combination of two factors. 
First, our new synthesis model predicts a higher total $\Gamma_{\rm HI}$ than HM12 at low redshift, by factor $\approx 2$ at $z=0$. 
Second, the \cite{2014ApJ...789L..32K} measurement is a factor $\approx3$ outlier in the opposite direction relative to subsequent $\Gamma_{\rm HI}$ measurements around the same redshift (also shown in Fig. \ref{fig:Gammas_lowz}). 
Together, these effects account for the factor of $\approx5$ discrepancy reported by \cite{2014ApJ...789L..32K}  between the HM12 synthesis model and their $\Gamma_{\rm HI}$ measurement. 
\cite{2015ApJ...811....3S} previously noted that assuming a higher escape fraction from star-forming galaxies could also reconcile observed galaxies and AGN with the low-redshift Ly$\alpha$ forest. 
Although this is possible, our analysis suggests that this is not required since AGN can account for most of the low-redshift ionizing background.

The HM12 synthesis model also significantly under-predicts the X-ray background measured by \citet[][]{2017ApJ...837...19C} 
using Chandra data at $\sim 2-7$ keV, by up to $\approx 30$\% \citep[see also][]{2006ApJ...645...95H}. 
However, HM12 show that their model are in better agreement with the HEAO-1 measurement from \cite{1999ApJ...520..124G} in this energy range. 
Figure \ref{fig:Jnu} shows that the difference in the X-ray spectrum between our synthesis model and HM12 increases in magnitude with increasing redshift. 
This is thus a regime where the two synthesis models can produce different results. 
The KS19 synthesis model is in better agreement with our new model at $\gtrsim 1$ keV. 
The FG09 model did not include X-rays from obscured AGN and therefore under-predicts the X-ray background at all redshifts. 

Another regime where current UVB models differ significantly is the energy range between 4 Ry (the HeII photoionization edge) and 1 keV. 
Figures \ref{fig:Gammas_full}, \ref{fig:Gammas_lowz}, and \ref{fig:Jnu} in particular show that our new UVB model predicts a specific intensity $J_{\nu}$ at energies near (but above) 4 Ry lower by a factor $\approx 2$ than the HM12 and KS19 models. 
Differences in this energy range arise because observational constraints on intrinsic AGN spectra and the IGM opacity in this regime are relatively poor, so the results are affected by different assumptions consistent with available data. 
In our model, the AGN spectral shape in this regime is set by connecting observational measurements in the UV ($\approx 600$~\AA) and in the soft X-rays ($\sim 1$ keV), while the HeII opacity is not directly constrained by observations but rather predicted by radiative transfer calculations (\S \ref{sec:rad_transp}). 
As can be seen in Figure \ref{fig:AGN_spectra}, our assumption that the AGN spectral template is a simple power-law from $600$~\AA~($\approx21$ eV) to 1 keV (distinct from the independently-constrained power-law immediately blueward of 912~\AA) implies a softening of the ionizing continuum at $600$~\AA.
For models calibrated to match the same HI ionization rate, this produces a lower and softer AGN flux near 4 Ry, relative to a single power-law extrapolation of the $F_{\nu} \propto \nu^{-1.7}$ continuum immediately blueward of 912~\AA~all the way to X-rays. 
This lower and softer HeII ionizing flux explains the systematic offsets relative to the HM12 and KS19 models, which assumed single power-law spectra from 912~\AA~to the soft X-rays.

\begin{figure*}
\begin{center}
\includegraphics[width=0.98\textwidth]{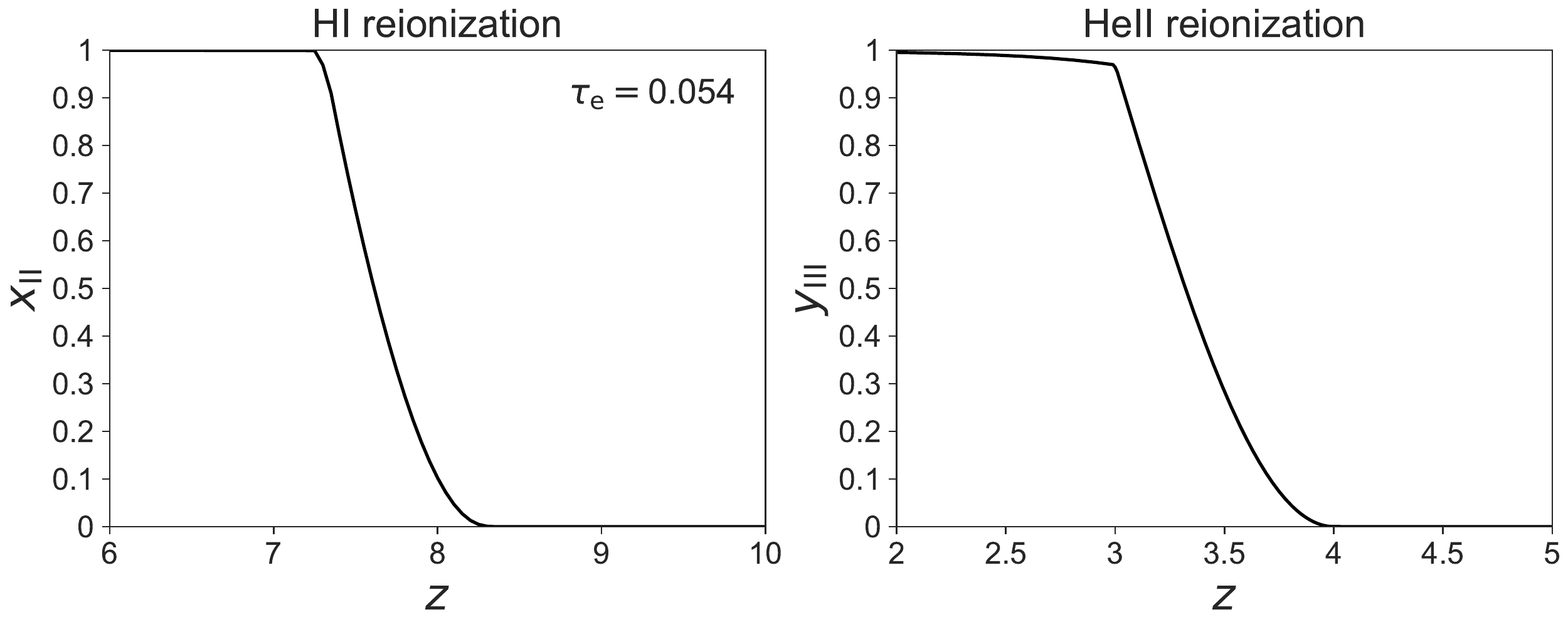}
\end{center}
\caption[]{Effective reionization history for our fiducial model (see \S \ref{sec:reionization_params} for parameters). 
The vertical axes show the volume-averaged HII and HeIII fractions. 
In this model, HI reionization ($z_{\rm rei,HI}=7.8$) is driven by star-forming galaxies and is calibrated to produce a CMB electron scattering optical depth that matches the best-fit value $\tau_{\rm e}=0.054$ measured by \cite{Planck-Collaboration:2018aa}. 
The timing of HeII reionization ($z_{\rm rei,HeII}=3.5$), driven by quasars, is in rough agreement with existing observational constraints from the HeII Ly$\alpha$ forest.
}
\label{fig:eff_reion_hist} 
\end{figure*}

\section{Effective reionization history}
\label{sec:reionization}
We now derive ``effective'' photoionization and photoheating rates designed to produce reionization histories with specified parameters. 
While reionization is expected to be highly inhomogeneous \citep[e.g.,][]{2001PhR...349..125B, 2006PhR...433..181F, 2011ASL.....4..228T}, most simulation codes do not follow the time-dependent radiative transfer necessary to model the inhomogeneities. Instead, most codes assume that the IGM is in photoionization equilibrium with a prescribed uniform homogeneous UVB. 
Our goal here is to provide homogeneous photoioionization rates constructed to produce a prescribed volume-averaged ionized fraction versus redshift under the assumption of photoionization equilibrium. 
We follow an approach similar to \cite{2017ApJ...837..106O}, who developed some ideas introduced in FG09.

\subsection{Effective photoionization rates}
\label{sec:effective_ionization}
We begin by describing how we modify photoionization rates. 
In photoionization equilibrium (neglecting collisional ionization),
\label{sec:reion}
\begin{equation}
\label{sec:hi_pei}
\Gamma_{\rm HI} n_{\rm HI} = C_{\rm HI} \alpha_{\rm HII}(T_{\rm IGM0,HI}) n_{\rm HII} n_{\rm e},
\end{equation}
where $\alpha_{\rm HII}$ is the recombination coefficient of HII into HI. 
The ionic number densities and IGM temperature $T_{\rm IGM0,HI}$ in equation (\ref{sec:hi_pei}) refer to values at the mean density of the Universe. 
The effects of clumping of intergalactic gas on photoionization balance are modeled through the clumping factor $C_{\rm HI}$. The HI subscripts here do not refer to the HI gas, but rather to the IGM temperature and clumping factor to use during HI reionization (recombinations require ionized gas). 
We assume a constant IGM temperature in constructing the effective photoionization rates, although in reality the IGM temperature changes during reionization events, e.g. owing to photoheating (\S \ref{sec:effective_heating}). 
We use the case A recombination coefficient because this is the limit normally used in cosmological simulation codes. 

Given a redshift-dependent hydrogen ionized fraction $x_{\rm II}(z)$, we can solve for the effective $\Gamma_{\rm HI}^{\rm rei}(z)$ that will produce the desired reionization history in photoionization equilibrium: 
\begin{equation}
\Gamma_{\rm HI}^{\rm rei}(z) = \frac{x_{\rm II}^{2}(z)}{1-x_{\rm II}(z)} (1 + \chi) C_{\rm HI} \alpha_{\rm HII}(T_{\rm IGM0,HI}) n_{\rm H}(z),
\end{equation}
where $\chi = Y/4X$. 
This result assumes that HeI is reionized simultaneously to HI (these two atoms have similar ionization potentials and can be simultaneously ionized by star-forming galaxies), but that HeII reionization is delayed until the rise of the AGN luminosity function. 
Using an expression analogous to equation (\ref{sec:hi_pei}) for photoionization equilibrium of HeI and assuming the same clumping factor, the assumption that hydrogen and helium are first ionized in concert ($x_{\rm II}=y_{\rm II}$) implies an effective HeI photoionization rate during reionization
\begin{equation}
\Gamma_{\rm HeI}^{\rm rei}(z) = \frac{\alpha_{\rm HeII}(T_{\rm IGM0,HI})}{\alpha_{\rm HII}(T_{\rm IGM0,HI})} \Gamma_{\rm HI}^{\rm rei}(z).
\end{equation}

Similarly, photoionization equilibrium for HeII in the two-state approximation that all He is either in HeII or HeIII, while H is fully ionized, implies 
\begin{equation}
\label{sec:heii_pei}
\Gamma_{\rm HeII} n_{\rm HeII} = C_{\rm HeII} \alpha_{\rm HeIII}(T_{\rm IGM0,HeII}) n_{\rm HeIII} n_{\rm e},
\end{equation}
where $T_{\rm IGM0,HeII}$ is the assumed IGM temperature at mean density during HeII reionization.
The effective HeII photonization rate is therefore
\begin{align}
\Gamma_{\rm HeII}^{\rm rei}(z) = \frac{y_{\rm III}(z)}{1-y_{\rm III}(z)} 
 & \left[1 + \chi (1+y_{\rm III}(z)) \right] \\ \notag
& ~~~~~\times C_{\rm HeII} \alpha_{\rm HeIII}(T_{\rm IGM0,HeII}) n_{\rm H}(z).
\end{align}

We parametrize reionization events using a redshift $z_{\rm rei,i}$ and a redshift width $\Delta z_{\rm rei,i}$. 
We define a smoothly varying function which tends to one for $z \ll z_{\rm rei,i}$ and to zero for $z \gg z_{\rm rei,i}$. 
We choose a function with compact support, which has the advantage of unambiguously defining redshifts at which reionization starts and ends:
\begin{equation}
\Theta(z;~z_{\rm rei,i},~\Delta z_{\rm rei,i}) = 1 + \sin
\left[
\frac{(z_{\rm rei,i}-z-\Delta z_{\rm rei,i})\pi}{4\Delta z_{\rm rei,i}}
\right],
\end{equation}
for $z \in [ z_{\rm rei,i} - \Delta z_{\rm rei,i},~z_{\rm rei,i} + \Delta z_{\rm rei,i}]$. 
Note that this corresponds to one quarter of a sine period, which we define this way so the ionization fraction evolves more slowly at the beginning of reionization and more rapidly toward the end as ionized bubbles overlap. Because of this asymmetry, the redshift of 50\% ionization is slightly below $z_{\rm rei,i}$. 
We use this function to model the evolving ionization fractions during HI and HeII reionization as
\begin{equation}
x_{\rm II}(z) = \Theta(z;~z_{\rm rei,HI},~\Delta z_{\rm rei,HI}) x_{\rm II}^{\rm eq}(z_{\rm rei,HI} - \Delta z_{\rm rei,HI})
\end{equation} 
and
\begin{equation}
y_{\rm III}(z) = \Theta(z;~z_{\rm rei,HeII},~\Delta z_{\rm rei,HeII}) y_{\rm III}^{\rm eq}(z_{\rm rei,HeII} - \Delta z_{\rm rei,HeII}),
\end{equation}
where $x_{\rm II}^{\rm eq}(z)$ and $y_{\rm III}^{\rm eq}(z)$ are the ionization fractions assuming photoionization equilibrium (eqs (\ref{sec:hi_pei}) and (\ref{sec:heii_pei})), solved for by using the post-reionization HI and HeII photoionization rates calculated in the homogeneous background approximation (to be explicit, we denote these post-reionization, equilibrium rates $\Gamma_{\rm i}^{\rm eq}$ below). 
This ensures continuity in the final effective photoionization rates:
\begin{align}
\Gamma_{\rm i}^{\rm eff}(z) = 
\left\{
\begin{array}{ll}
\Gamma_{\rm i}^{\rm eq}(z) & z \leq z_{\rm rei,i} - \Delta z_{\rm rei,i} \\
\Gamma_{\rm i}^{\rm rei,i}(z) & z > z_{\rm rei,i} - \Delta z_{\rm rei,i}.
\end{array}
\right.
\end{align}

For our effective reionization models, we use $\Delta z_{\rm rei,i}$ values that can be smaller than the actual duration of reionization events. 
This is because the equilibrium ionization fraction for a spatially-homogeneous photoionization rate depends on local density. 
If we used a large $\Delta z_{\rm rei,i}$, then low-density regions would be reionized significantly before high-density regions.
This would introduce large differences in reionization time tied to local density, which is not an accurate model for the propagation of large-scale ionizing fronts during reionization \citep[e.g.,][]{2004ApJ...613....1F, 2007MNRAS.377.1043M}. 
It would also introduce large differences between volume-weighted and mass-weighted ionization histories. 
For these reasons, we do not attempt to match the reionization history implied by our effective photoionization rates to the detailed redshift evolution of the HI ionized fraction constrained by observations other than the CMB optical depth \citep[e.g.,][]{2018ApJ...864..142D, 2019MNRAS.484.5094G, 2019MNRAS.489.2669M}. 
Instead, we force different gas densities to be reionized at roughly the same redshift by using relatively small $\Delta z_{\rm rei,i}$ values.
\begin{figure*}
\begin{center}
\includegraphics[width=0.98\textwidth]{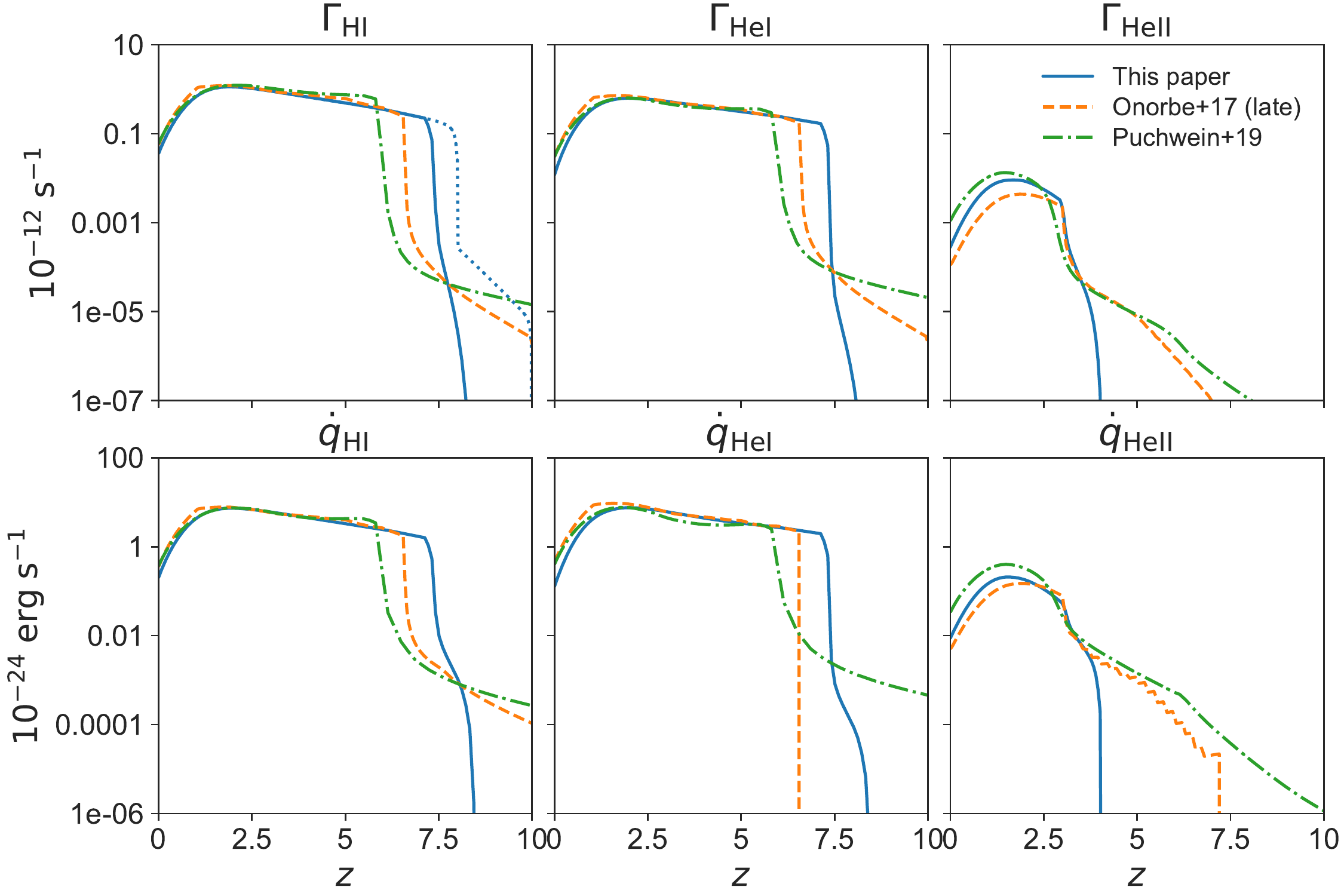}
\end{center}
\caption[]{
Effective photoionization (top) and photoheating (bottom) rates to model reionization in simulations that assume photoionization equilibrium with a homogeneous UVB. 
The solid curves correspond to our fiducial UVB model and Planck 2018 reionization history ($\tau_{\rm e}=0.054$; see \S \ref{sec:reionization_params} for reionization parameters). 
The sharp rises in the HI and HeI rates are set by the HI reionization redshift $z_{\rm rei,HI}=7.8$ and the sharp rises in the HeII rates are set by the HeII reionization redshift $z_{\rm rei,HeII}=3.5$. 
Following reionization events, the rates are direct integrals of the UVB intensity $J_{\nu}$ modeled assuming a homogeneous background (\S \ref{sec:models}). 
The dashed curves show rates for the `LateR' (late reionization) model from Onorbe et al. (2017). 
The dash-dotted curves show the fiducial model from Puchwein et al. (2019), which corresponds to a reionization optical depth $\tau_{\rm e}=0.065$ similar to the Planck 2015 result. 
Note that our model is constructed so that reionization events occur relatively rapidly to minimize differences in the reionization time of gas at different densities under the assumption of photoionization equilibrium.
}
\label{fig:TREECOOL_eff} 
\end{figure*}

\subsection{Effective photoheating rates}
\label{sec:effective_heating}
We can also improve the accuracy of the reionization treatment in simulations that assume a homogeneous UVB by defining effective photoheating rates during reionization. 
This is needed because the homogeneous approximation neglects optical depth effects which significantly affect photoheating during reionization \citep[e.g.][]{1999ApJ...520L..13A}.

Let us first focus on HI reionization. 
Assuming that photoheating due to reionization is perfectly coupled to the reionization process, we postulate that the rate of change of the IGM temperature during reionization can be approximated as
\begin{equation}
\label{eq:dTdt_HI_rei}
\left. \frac{dT}{dt} \right|_{\rm HI~rei} = \Delta T_{\rm HI} \frac{dx_{\rm II}}{dt},
\end{equation}
where $\Delta T_{\rm HI}$ is the total temperature increment due to reionization heating. 
Such an approach was suggested by \cite{2009ApJ...703.1416F} and implemented in more detail by \cite{2017ApJ...837..106O}. 
The HI photoheating rate is defined such that
\begin{align}
\dot{q}_{\rm HI} n_{\rm HI} &= \frac{{\rm heat~input~from~HI~photoionization}}{{\rm time } \times {\rm volume}},
\end{align} 
and analogous expressions define the HeI and HeII photoheating rates. 

As for the effective photoionization rates in the previous section, we can use this to define an effective total photoheating rate during HI reionization (arising as the sum of photoheating from the photoionization of HI and HeI):
\begin{align}
\left. \dot{q}_{\rm tot} \right|_{\rm HI~rei} n_{\rm HI} &= \left. \frac{d}{dt} \left( \frac{3 n_{\rm free} k_{\rm B} T}{2} \right) \right|_{\rm HI~rei},
\end{align}
where $n_{\rm free}$ is the total number of free particles that share the thermal energy. 
During HI reionization, $\Delta n_{\rm free}/n_{\rm free} \ll \Delta T/T$ since $n_{\rm free}$ changes only by order unity while the IGM temperature increases from $T\sim 10$ K to $T\sim 10^{4}$ K. 
This implies that $n_{\rm free} (dT/dt) \gg T (dn_{\rm free}/dt)$ and therefore
\begin{align}
\left. \dot{q}_{\rm tot} \right|_{\rm HI~rei} & \approx \frac{3}{2} \left( \frac{n_{\rm free}}{n_{\rm HI}} \right) \left. k_{\rm B} \frac{dT}{dt} \right|_{\rm HI~rei} \\ \notag
& = \frac{3}{2} \left( \frac{n_{\rm free}}{n_{\rm HI}} \right) k_{\rm B} \Delta T_{\rm HI} \frac{dx_{\rm II}}{dt} \\ \notag
& = \frac{3(1+\chi)}{2} \left(\frac{1+x_{\rm II}}{1-x_{\rm II}} \right) k_{\rm B} \Delta T_{\rm HI} \frac{dx_{\rm II}}{dt},
\end{align}
where we have used equation (\ref{eq:dTdt_HI_rei}) in the second step, and the same assumptions as before regarding the simultaneous reionization of HI and HeI in the third step.

Since we assume that HeI is reionized simultaneously to HI, we partition this total heating between the photionization of HI and HeI according to the relative number densities of H and He nuclei:
\begin{align}
\dot{q}_{\rm HI}^{\rm rei} &=  \frac{1}{1+\chi}   \left. \dot{q}_{\rm tot} \right|_{\rm HI~rei} \\ \notag
\dot{q}_{\rm HeI}^{\rm rei}&= \frac{\chi}{1+\chi} \left. \dot{q}_{\rm tot} \right|_{\rm HI~rei}.
\end{align}

Using analogous approximations as for HI reionization but now applied to HeII reionization, which we assume proceed after HI/HeI reionization has completed, we can derive a similar effective photoheating rate:
\begin{align}
\left. \dot{q}_{\rm HeII}^{\rm rei} \right|_{\rm HeII~rei} & \approx \frac{3}{2} \left( \frac{n_{\rm free}}{n_{\rm HeII}} \right) \left. k_{\rm B} \frac{dT}{dt} \right|_{\rm HeII~rei} \\ \notag
& = \frac{3}{2} \left( \frac{n_{\rm free}}{n_{\rm HeII}} \right) k_{\rm B} \Delta T_{\rm HeII} \frac{dy_{\rm III}}{dt} \\ \notag
& = \frac{3[2+\chi(2+y_{\rm III})]}{2\chi (1 - y_{\rm III})}
k_{\rm B} \Delta T_{\rm HeII} \frac{dy_{\rm III}}{dt}.
\end{align}

\subsection{Reionization history parameters}
\label{sec:reionization_params}
The parameters of the reionization history are the subject of active observational research. 
It is also useful for theoretical studies to explore the effects of different reionization histories, e.g. on the properties of dwarf galaxies in cosmological simulations (e.g. Ben\'itez-Llambay et al. 2015; Garrison-Kimmel et al., in prep.)\nocite{2015MNRAS.450.4207B}. 
Our approach is therefore to provide a UVB model with a fiducial reionization history motivated by current constraints and to also make available, in electronic form, data for other reionization parameters.

An important calibration for our UVB models is that they should produce a reionization history consistent with the electron scattering optical depth to the surface of last scattering measured from CMB observations. 
This optical depth is given by
\begin{equation}
\tau_{\rm e} = \int_{0}^{\infty} dz \frac{dl}{dz} n_{\rm e}(z) \sigma_{\rm T},
\end{equation}
where $\sigma_{\rm T}$ is the Thomson cross section.

For our fiducial reionization model, we assume the following parameters for HI reionization:
\begin{align}
z_{\rm rei,HI} &= 7.8 \\ \notag
\Delta z_{\rm rei,HI} &= 0.5 \\ \notag
C_{\rm HI} &= 3 \\ \notag
T_{\rm IGM0,HI} &= 10,000~{\rm K} \\ \notag
\Delta T_{\rm HI} &= 20,000~{\rm K}
\end{align} 
and the following parameters for HeII reionization:
\begin{align}
z_{\rm rei,HeII} &= 3.5 \\ \notag
\Delta z_{\rm rei,HeII} &= 0.5 \\ \notag
C_{\rm HeII} &= 3 \\ \notag
T_{\rm IGM0,HeII} &= 10,000~{\rm K} \\ \notag
\Delta T_{\rm HeII} &= 15,000~{\rm K}.
\end{align} 
Figure \ref{fig:eff_reion_hist} shows the resulting ionization fractions of HI and HeII, assuming photoionization equilibrium. 
The electron scattering optical depth for this model is $\tau_{\rm e}=0.054$, equal to the best fit to the Planck 2018 CMB data. 

The other reionization parameters were chosen as follows. 
The redshift of HeII reionization, $z_{\rm rei,HeII} = 3.5$, is consistent with simulations of HeII reionization that match the observationally-inferred HeII photoionization rate at $z\sim2.5-3$ \citep[e.g.,][]{2009ApJ...694..842M, 2018arXiv180805247W} and is also broadly consistent with inferences from the evolution of the IGM temperature \citep[e.g.,][]{2010ApJ...718..199L, 2011MNRAS.410.1096B, 2019ApJ...872...13W}. 
We use approximate clumping factors motivated by a combination of numerical simulations and analytic arguments for the gas in which IGM recombinations take place \citep[e.g.,][]{2011ApJ...743...82M, 2012MNRAS.427.2464F, 2014MNRAS.443.2722J, 2014ApJ...787..146K, 2015ApJ...810..154K}. 
The mean-density IGM temperatures to use here should be representative of the redshift when most recombinations take place during reionization, i.e. around the reionization mid-point. 
We use a rough value of $T_{\rm IGM0,HI} = T_{\rm IGM0,HeII} = 10^{4}$ K; since the recombination coefficients scale as $\propto T^{-0.6}$ in the relevant temperature range, the exact choice has only a modest impact on the results.
For HI reionization heating, we use a value $\Delta T_{\rm HI} = 20,000$ K representative of standard galaxy-driven reionization \citep[e.g.][]{2012MNRAS.426.1349M} while for HeII reionization heating we use a value $\Delta T_{\rm HeII} = 15,000$ K appropriate for reionization by quasars with an extreme-UV spectrum $f_{\nu} \propto \nu^{-1.7}$ (e.g., McQuinn et al. 2019; FG09).\nocite{2009ApJ...694..842M} 

The effective photoionization and photoheating rates for this fiducial reionization model are plotted in Figure \ref{fig:TREECOOL_eff} and tabulated in Appendix \ref{sec:rates_appendix}. 
The figure also shows effective reionization rates from \cite{2017ApJ...837..106O} and \cite{2019MNRAS.485...47P}.

\section{Discussion and conclusions}
\label{sec:discussion}
In this paper, we updated the \cite{2009ApJ...703.1416F} cosmic UVB model. 
Our goal was to incorporate several new observational constraints on the galaxy and AGN populations that putatively dominate the background, an improved model of the IGM opacity, as well as some new modeling elements. 
Among the new modeling elements, 
we used the BPASS spectral model for star-forming galaxies including binary stars, a combination of obscured and unobscured AGN (necessary to simultaneously match the empirical UV and X-ray backgrounds), and the HeII Lyman-series sawtooth feature \citep[][]{2009ApJ...693L.100M}. 
We also computed effective photoionization and photoheating rates to produce a reionization history consistent with the latest electron scattering optical depth from Planck \citep{Planck-Collaboration:2018aa} and constraints on HeII reionization from quasar absorption spectra in simulations that assume photoionization equilibrium with a homogeneous background. 

For our new model, which provides an overall good match to a range of observational constraints, we find the following:
\begin{enumerate}
\item The HI ionizing background, quantified by the photoionization rate $\Gamma_{\rm HI}$, is dominated by AGN at $z\lesssim3$ and by star-forming galaxies at $z\gtrsim3$. 
This is the case even though the non-ionizing background at rest-frame wavelength UV $1500$~\AA~is dominated by star-forming galaxies at all redshifts. 
\item AGN can explain the entire HI ionizing background at $z<0.5$ inferred by recent modeling of the Ly$\alpha$ forest, 
assuming a standard spectral energy distribution template and unity escape fraction for AGN. 
Our UVB model thus suggests that there is no ``photon underproduction crisis'' \citep[cf.,][]{2014ApJ...789L..32K}.
\item The fact that AGN can explain the low-redshift HI ionizing background but fall increasingly short of explaining the Ly$\alpha$ forest transmission at $z>3$ due to the steeply declining AGN luminosity function suggests that the population-averaged escape fraction of ionizing photons from star-forming galaxies increases strongly from $z\sim0$ to $z \gtrsim6$. This could be due to powerful stellar feedback clearing large ``holes'' in the ISM of early galaxies. 
\item Our UVB synthesis model matches the total HI photoionization rate inferred from the Ly$\alpha$ forest for an effective (UV emissivity-weighted) absolute escape fraction of 1\% for $z=3$ galaxies. 
This is substantially lower than the escape fractions $\approx5-10$\% implied by recent direct measurements of escaping Lyman continuum photons \citep[e.g.,][]{2018ApJ...869..123S, 2019ApJ...878...87F}. This may indicate that escape fraction trends observed in current direct measurement samples cannot be extrapolated to fainter galaxies.
\item The non-ionizing part of the low-redshift UVB spectrum inferred from cross-correlating GALEX observations with SDSS spectroscopic objects \citep[][]{2019ApJ...877..150C} is in excellent agreement with the UVB spectrum predicted using the BPASS stellar template. 
\item The low-redshift HI ionizing background and the local X-ray background can be simultaneously explained by an AGN population that includes both obscured and unobscured sources and are consistent with a large AGN obscured fraction $f_{\rm obsc}\approx 0.75$.  
\item As in previous models, the HeII ionizing background is dominated by AGN at all redshifts due to the strong spectral break at the HeII photoionization edge in stellar spectra. 
While star-forming galaxies drive HI reionization, AGN drive HeII reionization at lower redshifts. 
\end{enumerate}

One motivation for updating our UVB model was to better understand the uncertainties in available UVB models. 
We find that different recent synthesis models (e.g., HM12, KS19, P19, and this work) are qualitatively similar in their predicted spectra and redshift evolution. 
However, important quantitative differences remain. 
Relative to HM12, our new model provides a better match to recent measurements of the low-redshift Ly$\alpha$ forest (higher $\Gamma_{\rm HI}$) as well as to the local X-ray background determined by Chandra. 
Furthermore, the differences in X-ray predictions increase with increasing redshift. 
Different models also differ notably in their predictions in the energy range between the HeII photoionization edge (4 Ry) and $\approx 1$ keV, due to relatively poor observational constraints on the intrinsic AGN spectrum and IGM opacity in this regime. 
As for the modeling of reionization, to our knowledge this work presents the first full UVB spectral synthesis model to include effective photoionization and photoheating rates that match the best-fit Planck 2018 electron scattering optical depth, corresponding to a relatively late redshift of HI reionization $z_{\rm rei,HI}\sim7.8$. 

We plan to release model updates as improved empirical constraints become available.\footnote{See http://galaxies.northwestern.edu/uvb for the electronic data.} 
 
\section*{Acknowledgments}
We acknowledge Jos\'e O\~norbe, Jonathan Stern, and Luke Zoltan Kelley for useful discussions and comments. 
We thank Andrew Wetzel, Nick Frontiere and JD Emberson for testing our new model UVB in hydrodynamic simulations. 
We are grateful to Rychard Bouwens for providing galaxy luminosity function measurements ahead of publication, and to Jacob Shen for providing updated AGN luminosity function and spectral models also before publication. 
Detailed comments by an anonymous referee substantially improved the paper. 
This work was supported by NSF through grants AST-1517491, AST-1715216, and CAREER award AST-1652522; by NASA through grants NNX15AB22G and 17-ATP17-0067; by STScI through grants HST-GO-14681.011, HST-GO-14268.022-A, and HST-AR-14293.001-A; and by a Cottrell Scholar Award from the Research Corporation for Science Advancement.

\appendix

\section{Ionizing mean free paths}
\label{sec:fesc_mfp_deg}
UVB models depend on assumptions about source populations as well as the transfer of radiation through intervening gas. 
A key parameter is the mean free path (m.f.p.) of ionizing photons. 
This is most clearly seen by considering the ``local source approximation,'' which is valid at redshifts and frequencies such that the m.f.p. in the IGM, $\Delta l$, is much shorter than the Hubble length. 
In this limit, the UVB intensity at any location can be thought of as arising from the emissivity integrated over all sources within a radius $\Delta l$, i.e. 
\begin{align}
J_{\nu} \approx \frac{1}{4\pi} \epsilon_{\nu} \Delta l(\nu)
\end{align}
(see equation D1 in FG09). 
 \begin{figure*}[ht]
\begin{center}
\includegraphics[width=0.99\textwidth]{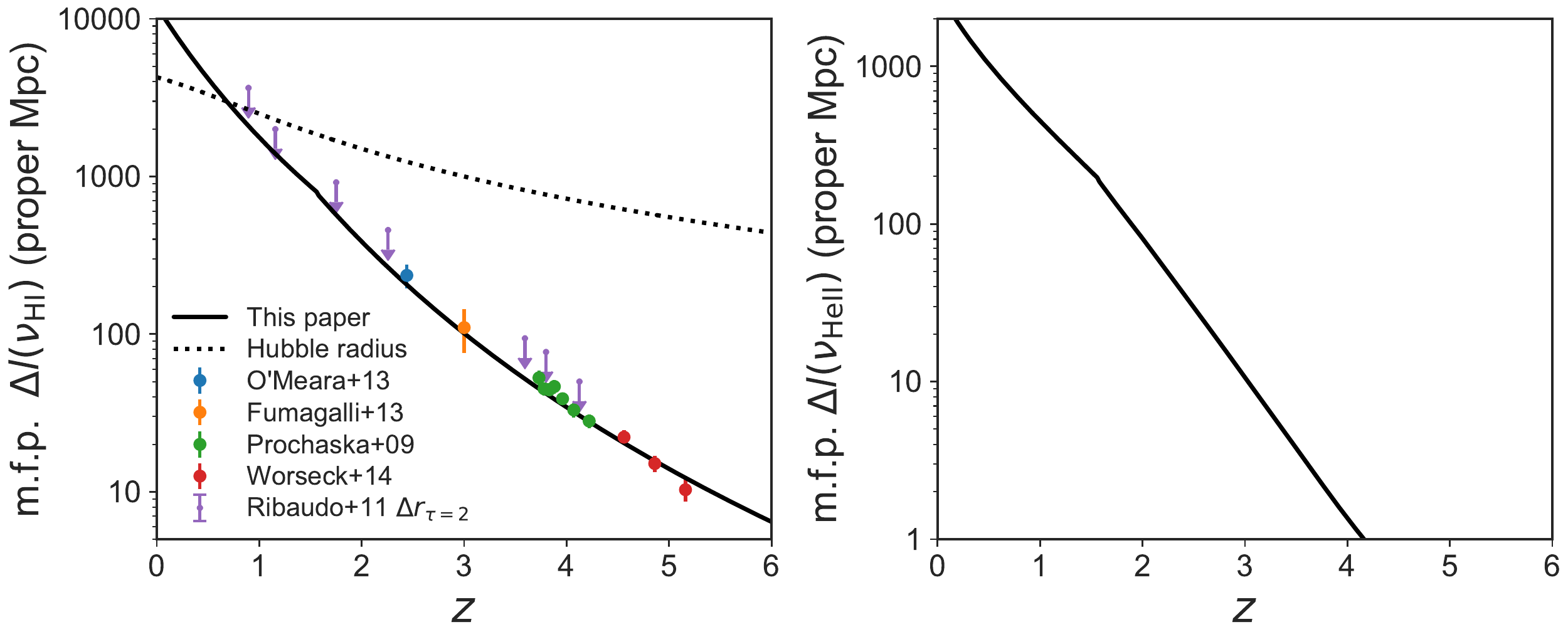}
\end{center}
\caption{\emph{Left:} HI ionizing mean free path implied by the column density distribution adopted in this paper (solid curve) compared to observational measurements of the m.f.p.. 
The purple symbols show the mean spacing between HI absorbers with optical depth $\tau=2$ at the Lyman limit and correspond to upper limits on the m.f.p.. 
The dotted curve shows the Hubble length $c/H(z)$; this length becomes important for limiting the HI ionizing rate when it becomes shorter than the m.f.p., i.e. at $z\lesssim1$. 
\emph{Right:} The HeII ionizing m.f.p. for our fiducial UVB model.}
\label{fig:mfp} 
\end{figure*}

Since photoionization rates scale linearly with $J_{\nu}$ we have, for example, 
\begin{align}
\label{eq:fesc_mfp_deg}
\Gamma_{\rm HI} \propto \frac{1}{4\pi} \epsilon_{\nu 1500}^{\star,\rm prop} 
f_{\rm esc}^{\star} \Delta l(\nu_{\rm HI})
\end{align}
in the high-redshift limit in which star-forming galaxies dominate the HI ionizing background. 
Equation (\ref{eq:fesc_mfp_deg}) highlights the degeneracy between the m.f.p. model and the escape fraction inferred by comparing the emissivity obtained by integrating over the galaxy luminosity function ($\epsilon_{\nu 1500}^{\star}$) and the total IGM photoionization rate measured using the Ly$\alpha$ forest ($\Gamma_{\rm HI}$). 

The m.f.p. corresponds to the length over which the increment of effective optical depth $\bar{\tau}$ is unity. 
It can therefore be obtained by evaluating the following derivative:
\begin{equation}
\Delta l(\nu,~z) = \left( \frac{d\bar{\tau}}{dl} \right)^{-1}(\nu,~z),
\end{equation}
where $\bar{\tau}$ is given by equation (\ref{taueff poisson expression}). 
The left hand side of Figure \ref{fig:mfp} shows the HI ionizing m.f.p. (at $\nu_{\rm HI}$) corresponding to the HI column density distribution adopted in our UVB model (see \S \ref{sec:rad_transp}). 
This model m.f.p. is compared to observational measurements from \cite{2013ApJ...765..137O}, \cite{2013ApJ...775...78F}, \cite{2009ApJ...705L.113P}, and \cite{2014MNRAS.445.1745W}. 
We also show on this panel the mean distance between HI absorbers with optical depth $\tau=2$ at the Lyman limit from \cite{2011ApJ...736...42R}. 
This is an upper limit to the HI ionizing m.f.p. because it neglects opacity from absorbers with $\tau<2$. 
Overall, our model is in excellent agreement with observational constraints on the HI ionizing m.f.p.. 
The dotted curve shows the Hubble length $c / H(z)$. 
Comparing the m.f.p. with the Hubble length indicates that the IGM photoionization rate is limited by the Hubble volume at $z\lesssim1$, and therefore that the local source approximation is only valid at $z \gtrsim 1$. 

The HeII ionizing m.f.p. (at $\nu_{\rm HeII}$) cannot be computed simply by integrating over the assumed HI column density distribution since it depends on the ratio $\eta = N_{\rm HeII}/N_{\rm HI}$, which itself is a function of the UVB. 
The HeII ionizing m.f.p. can however been inferred from a full calculation of the UVB spectrum vs. redshift; it is shown on the right hand side of Figure \ref{fig:mfp} for the fiducial UVB model in this paper.
\begin{figure*}
\begin{center}
\includegraphics[width=0.99\textwidth]{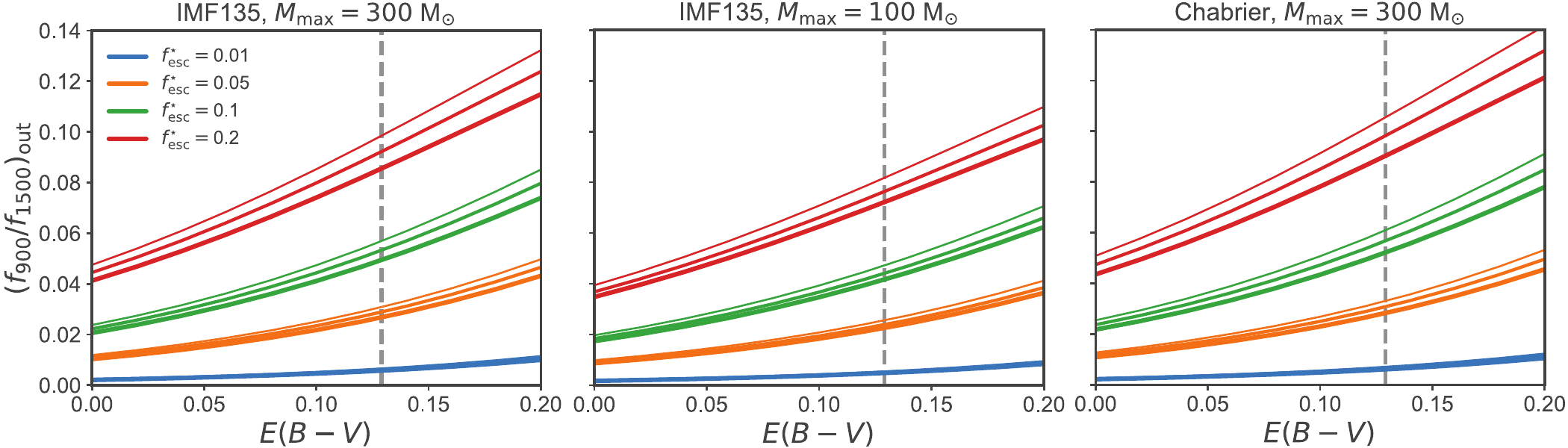}
\end{center}
\caption{Exploration of how the ratio of the flux escaping from galaxies at 900~\AA~vs. 1500~\AA~depends on the assumed IMF, dust reddening $E(B-V)$, and absolute escape fraction $f_{\rm esc}^{\star}$ defined in the holes model (eq. \ref{eq:holes_model}). 
\emph{Left:} Default IMF used in the UVB model (see \S \ref{sec:stellar_spec}). \emph{Middle:} Same functional form but with a maximum stellar mass of 100~M$_{\odot}$ instead 300~M$_{\odot}$. 
\emph{Right:} A Chabrier (2003) IMF with a maximum mass of 300~M$_{\odot}$. 
In each panel and for each $f_{\rm esc}^{\star}$, three curves of increasing thickness correspond to stellar metallicities $Z_{\star}=0.1,~0.3,~1$ $Z_{\odot}$.
All curves assume an intrinsic stellar population spectrum including binaries predicted by BPASS. 
}
\label{fig:f900f1500_vs_IMF} 
\end{figure*}

\section{Stellar spectra and escape fraction}
\label{sec:stellar_fesc}
Assuming a simple power-law spectrum $J_{\nu} = J_{\nu_{\rm HI}} (\nu / \nu_{\rm HI})^{-\alpha_{\rm HI}}$ blueward of the Lyman limit, the local source approximation implies
\begin{equation}
\label{eq:ls_pl}
\Gamma_{\rm HI} \approx \frac{\sigma_{\rm HI} \epsilon_{\nu 912}^{\rm \star,prop} \Delta l(\nu_{\rm HI}) }{h(\alpha_{\rm HI}+3)}
\end{equation}
\citep[eq. 25 in][]{2008ApJ...688...85F}. 
Consider a high redshift where star-forming galaxies dominate the HI ionizing background,
\begin{equation}
\label{eq:eps_ratio}
\frac{\epsilon_{\nu 912}^{\rm \star,prop}}{\epsilon_{\nu 1500}^{\rm \star,prop}} \approx \frac{S_{\nu 912}}{S_{\nu 1500}},
\end{equation}
where $S_{\nu}$ is the spectral template for stellar radiation escaping galaxies into the IGM (eq. \ref{eq:holes_model}). 
In the limit $f_{\rm esc}^{\star} \ll 1$ and $f_{\rm esc}^{\star} \ll f_{\rm dust}^{1500}$, where $f_{\rm dust}({\rm 1500~\AA}) \equiv 10^{-0.4E(B-V)k(1500~{\rm \AA})}$, 
the holes model implies the simple scaling
\begin{equation}
\label{eq:S_ratio}
\left( \frac{f_{900}}{f_{1500}} \right)_{\rm out} 
\equiv
\frac{S_{\nu 912}}{S_{\nu 1500}} \approx \left( \frac{S_{\nu 912}^{\rm intr}}{S_{\nu 1500}^{\rm intr}} \right) 
\frac{f_{\rm esc}^{\star}}{f_{\rm dust}^{1500}},
\end{equation}
where we have introduced the notation $(f_{900}/f_{1500})_{\rm out}$ to connect to observational studies which measure the ratio of the flux escaping galaxies at $\approx 900$~\AA~to the flux escaping at $\approx 1500$~\AA. 
This ratio is more directly constrained by observations than the absolute fraction of ionizing photons that escape galaxies, since the estimating the latter requires a model for the intrinsic ionizing flux of a stellar population \citep[for a discussion of different definitions of the escape fraction, see][]{2018ApJ...869..123S}. 
In our numerical evaluations of the fluxes from BPASS spectra, we average $S_{\nu}$ over the wavelength windows 1460-1510~\AA~and 890-910~\AA~to avoid local fluctuations due to emission and absorption lines (including an SiII absorption line at 1526~\AA). 
The right-hand side expresses this ratio in terms of the ratio of intrinsic specific fluxes at these rest wavelengths, and the effects of the escape fraction of ionizing photons and dust attenuation. 
Only the dust attenuation at 1500~\AA~appears in this expression because escaping ionizing photons are not affected by dust in the holes model (the holes are clear of both ionizing and dust opacity). 
Combining equations (\ref{eq:ls_pl})-(\ref{eq:S_ratio}), we find:
\begin{align}
\label{eq:GHI_vs_fesc}
\Gamma_{\rm HI} & \approx \frac{\sigma_{\rm HI}}{h(\alpha_{\rm HI}+3)} \left( \frac{S_{\nu 912}^{\rm intr}}{S_{\nu 1500}^{\rm intr}} \right) \left( \frac{f_{\rm esc}^{\star}}{f_{\rm dust}^{1500}} \right) \epsilon_{\nu 1500}^{\rm \star,prop} \Delta l(\nu_{\rm HI}) \\ \notag
& \approx \frac{\sigma_{\rm HI}}{h(\alpha_{\rm HI}+3)} 
\left( \frac{f_{900}}{f_{1500}} \right)_{\rm out} 
 \epsilon_{\nu 1500}^{\rm \star,prop} \Delta l(\nu_{\rm HI}).
\end{align}
This is the more complete expression corresponding to the scaling in equation (\ref{eq:fesc_mfp_deg}) above. 

Equation (\ref{eq:GHI_vs_fesc}) shows that, for assumed $\alpha_{\rm HI}$ and $\Delta l(\nu_{\rm HI})$, the combination of a Ly$\alpha$ forest $\Gamma_{\rm HI}$  measurement and the UV luminosity function (which yields $\epsilon_{\nu 1500}^{\rm \star,prop}$) determines what the effective $(f_{900}/f_{1500})_{\rm out}$ must be. 
In Figure \ref{fig:f900f1500_vs_IMF}, we plot $(f_{900}/f_{1500})_{\rm out}$ vs. $E(B-V)$ for different assumed IMFs, stellar metallicities $Z_{\star}$, and absolute escape fractions $f_{\rm esc}^{\star}$. 
These curves can be used to gauge how the inferred absolute escape fraction depends on assumed parameters.
All curves assume BPASS v2.2.1. 
We find relatively small differences ($\sim 10\%$) when either the stellar IMF or metallicity is varied. 
Going from the fiducial $E(B-V)=0.129$ to no dust extinction can reduce $(f_{900}/f_{1500})_{\rm out}$ at given $f_{\rm esc}^{\star}$ by a factor $\sim2$, i.e. imply a $\sim2\times$ larger $f_{\rm esc}^{\star}$ for a given $(f_{900}/f_{1500})_{\rm out}$. 
Unless our understanding of how ionizing photons propagate in the IGM is affected by much larger errors than is presently appreciated \citep[e.g.,][]{2014MNRAS.438..476P}, we do not expect that the $\Delta l(\nu_{\rm HI})$ term or the prefactor involving $\alpha_{\rm HI}$ in equation (\ref{eq:GHI_vs_fesc}) to be off by more than tens of percent. 
Thus, overall, it appears difficult to reconcile the effective $f_{\rm esc}^{\star,z=3}\approx1$\% absolute escape fraction implied by our UVB synthesis model \citep[and others;][]{2012ApJ...746..125H, 2019MNRAS.485...47P, 2019MNRAS.484.4174K} with recent measurements from the direct detection of escaping Lyman continuum photons implying $\approx5-10$\% escape fractions at $z\approx3$ \citep[][]{2018ApJ...869..123S, 2019ApJ...878...87F}, other than if faint galaxies not represented in direct studies have much smaller escape fractions (see \S \ref{sec:stellar_spec}). 

\section{AGN luminosity functions}
\label{sec:AGN_appendix}
\begin{figure}
\begin{center}
\includegraphics[width=0.47\textwidth]{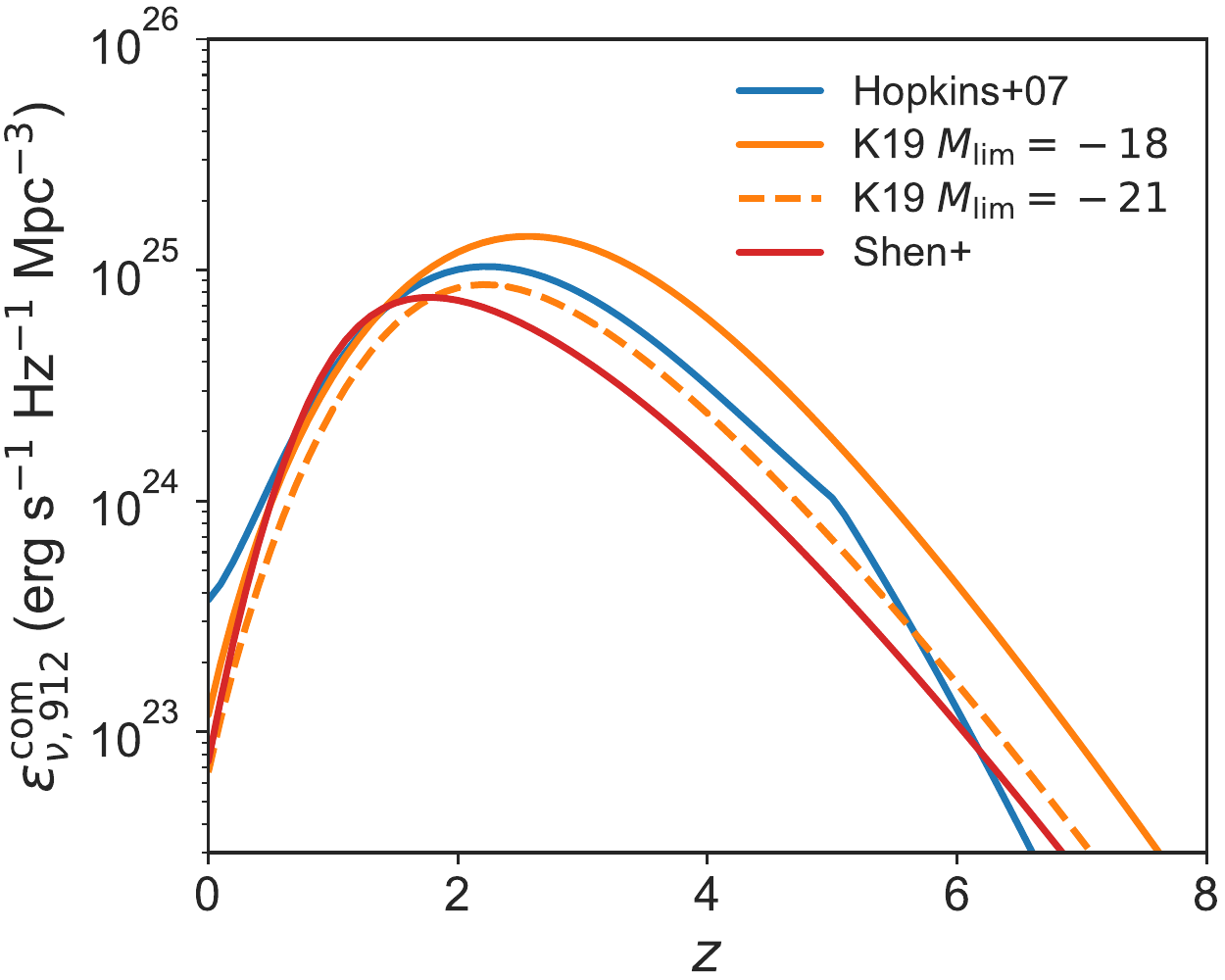}
\end{center}
\caption{
Comparison of ionizing emissivities (specific emissivities at 912~\AA) implied by different AGN luminosity functions. 
The solid blue curve shows the ionizing emissivity implied by the $B-$band luminosity function from Hopkins et al. (2007), which was used in FG09. 
The orange curves (solid and dashed) show the ionizing emissivities from Kulkarni et al. (2019) based on a newer compilation of UV luminosity functions, integrated down to limiting 1450~\AA~UV magnitudes of -18 and -21, respectively. 
The solid red curve shows the ionizing emissivity 
for the Shen et al. (2020) update of the Hopkins et al. (2007) bolometric luminosity function (see eq. \ref{eq:shen_eps_fit}). 
The emissivities implied by the Hopkins et al. and Shen et al. luminosity functions converge well with limiting magnitude, but the emissivity implied by the Kulkarni et al. luminosity function is more sensitive to the assumed limiting magnitude (see Fig. \ref{fig:QLF_completeness_diff}). 
To evaluate the specific emissivity at 912~\AA~from the Hopkins et al. 
luminosity function, we use AGN spectral template described in \S \ref{sec:agn_spec}; for the Kulkarni et al. luminosity function, we use the fits provided by the authors for the ionizing emissivities corresponding to different limiting magnitudes. 
We assume an escape fraction of unity for ionizing photons in each case. 
}
\label{fig:eps912_vs_z_AGN} 
\end{figure}

\begin{figure*}
\begin{center}
\mbox{
\includegraphics[width=0.49\textwidth]{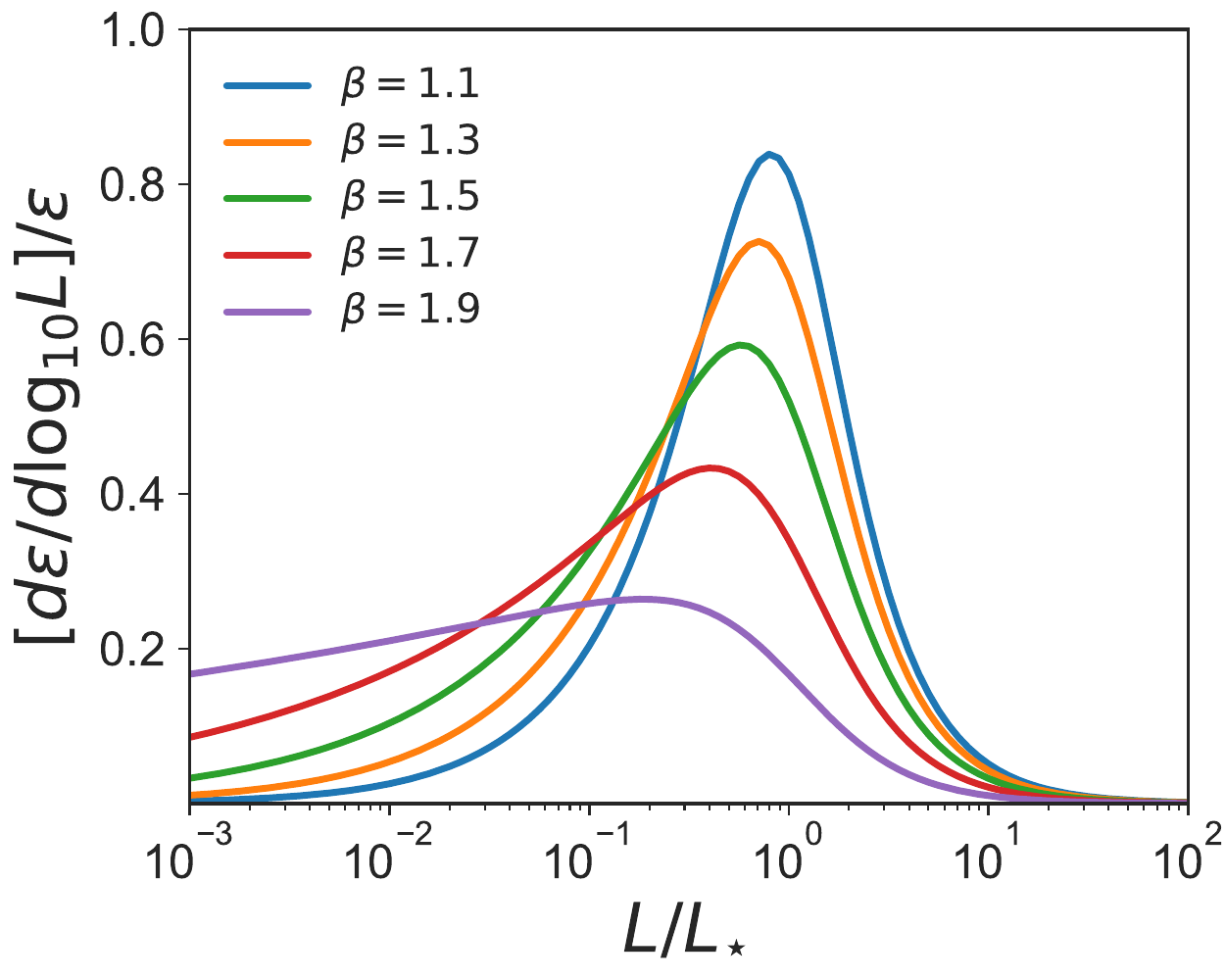}
\includegraphics[width=0.49\textwidth]{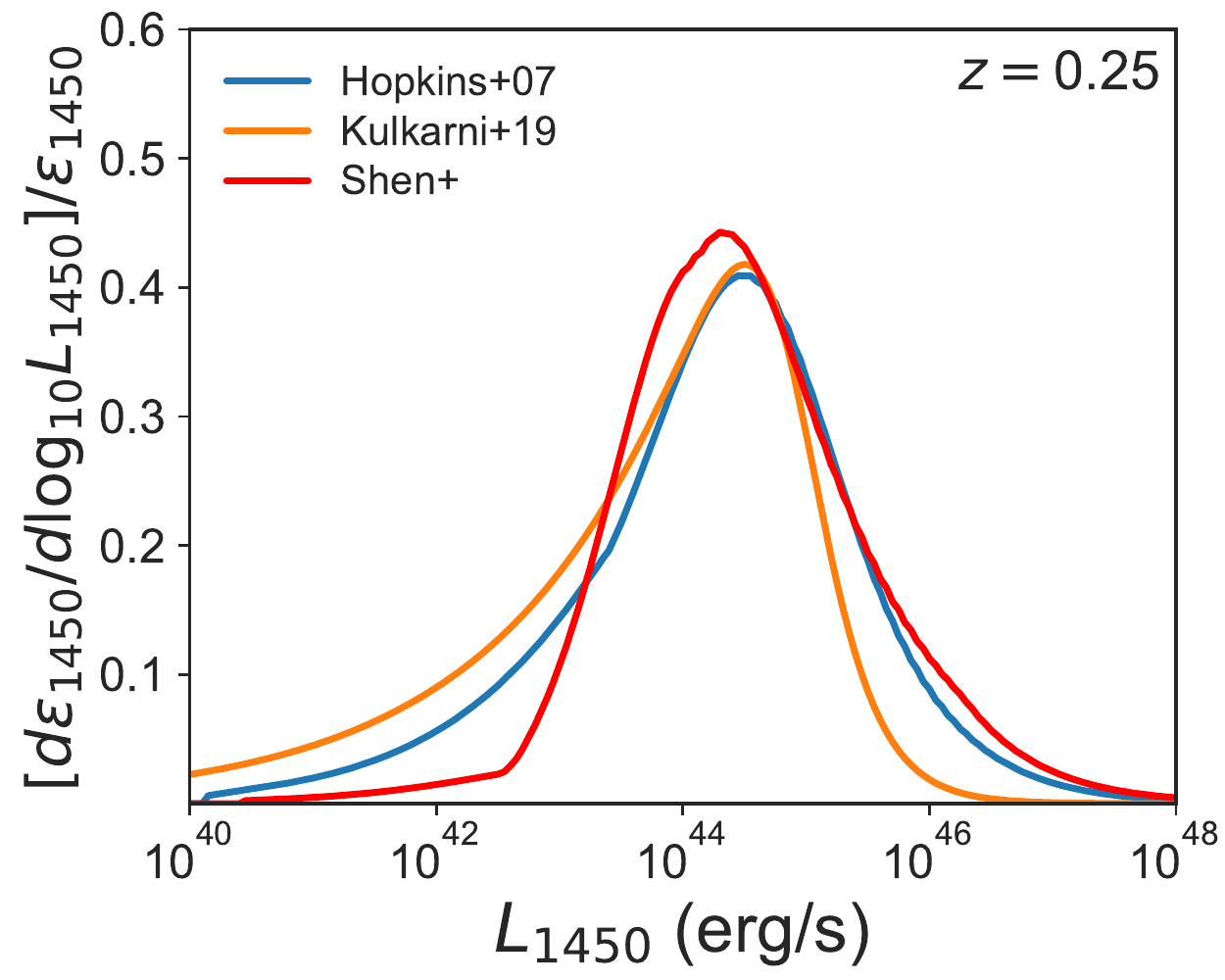}
}
\end{center}
\caption{
Differential contributions to the total emissivity. 
\emph{Left:} Different power-law luminosity function models, with varying faint-end slope $\beta$ and fixed bright-end slope $\alpha=3.5$, expressed in terms of luminosity normalized to the break luminosity $L_{\star}$. 
Steeper faint-end slopes cause the total emissivity to depend sensitively how far the integral is extrapolated below the luminosities of well sampled AGN. 
\emph{Right:} Similar but for rest-UV 1450~\AA~AGN~luminosity functions from Hopkins et al. (2007), Kulkarni et al. (2019), and Shen et al. (2020) at $z=0.25$. 
For the Hopkins et al. curve, we first evaluate the $B-$band (4400~\AA) luminosity function (as in FG09) and convert to 1450~\AA~using the spectral template in \S \ref{sec:agn_spec}. 
In this figure, $L_{1450}= \nu L_{\nu} |_{\rm 1450~\AA}$. 
}
\label{fig:QLF_completeness_diff} 
\end{figure*}

In Figure \ref{fig:eps912_vs_z_AGN} we compare the ionizing emissivities vs. redshift implied by different AGN luminosity functions (Hopkins et al. 2007 [used in FG09]; Kulkarni et al. 2019; and Shen et al. 2020 [used in this work]). 
The emissivities implied by the Hopkins et al. and Shen et al. luminosity functions converge well with limiting magnitude (see Fig. \ref{fig:QLF_completeness_diff}), so the curves shown correspond to total emissivities. 
For the Kulkarni et al. luminosity function, we show the results of integrating town to limiting rest 1450~\AA~UV magnitudes of -18 and -21, respectively; the curves are significantly different owing to the faint-end slope of the Kulkarni et al. luminosity function. 
We assume an escape fraction of unity for ionizing photons in each case. 

A significant difference between the \cite{2007ApJ...654..731H} luminosity function and the more recent \cite{2019MNRAS.488.1035K} and \cite{2020arXiv200102696S} determinations is the emissivity implied at $z<0.5$. 
Namely, the more recent luminosity functions imply emissivities that decline much more rapidly from $z\approx 0.5$ to $z=0$, resulting in a factor $\sim3$ difference by $z=0$. 
This difference is important because this is a redshift regime in which our fiducial model implies that AGN dominate the ionizing emissivity over star-forming galaxies, so the AGN ionizing emissivity primarily determines whether we can explain the ionization rate implied by the Ly$\alpha$ at low redshift (see Figure \ref{fig:Gammas_lowz}). 
Our results indicate that AGN can indeed provide the majority of the ionizing photons required by the low-redshift Ly$\alpha$ forest, but only if the AGN that contribute significantly the ionizing background have an escape fraction near unity, as we have assumed. 

At high redshift ($z \gtrsim 3$), the different luminosity functions also predict integrated ionizing emissivities that can differ by a factor $\sim3$. 
The largest difference is between the \cite{2020arXiv200102696S} luminosity function and the \cite{2019MNRAS.488.1035K} luminosity function integrated down to a limiting magnitude of -18. 
On the other hand, the \cite{2019MNRAS.488.1035K} luminosity function integrated to a limiting magnitude of -21, and the \cite{2007ApJ...654..731H} and \cite{2020arXiv200102696S} luminosity functions integrated over all luminosities predict emissivities within a factor of 2 of each other from $z\sim0.5$ to $z\approx 6$. 
Since star-forming galaxies increasingly dominate the HI ionizing background at high redshift, the relatively large differences between AGN luminosity functions have only modest effects on the total HI ionizing background, although the effects can be larger at higher energies.

We now show how the shape of the AGN luminosity function affects which AGN contribute the most to the ionizing emissivity. 
We consider an arbitrary waveband and drop waveband-specific subscripts to simplify the notation. 
We define $d\phi/dL$ as the number density of sources with luminosity in the interval $dL$ and $d\phi/dM$ as the number density of sources with absolute magnitude in the interval $dM$. 
It is common to use a double power-law function to characterize the AGN luminosity function. 
In terms of magnitudes, we define:
\begin{equation}
\frac{d\phi}{dM} = \frac{\phi_{\star}}{10^{0.4(\alpha+1)(M-M_{\star})} + 10^{0.4(\beta+1)(M-M_{\star})}},
\end{equation}
where $\phi_{\star}$ is the amplitude, $M_{\star}$ is the break magnitude, $\alpha$ is the bright-end slope, and $\beta$ is the faint-end slope. 
Using the usual conversion between AB magnitude and specific luminosity,
\begin{equation}
L_{\nu} = 4\pi (10~{\rm pc~cm^{-1}})^{2}~10^{-0.4(M+48.60)}~{\rm erg~s^{-1}~Hz^{-1}}
\end{equation}
\citep[][]{1983ApJ...266..713O}, the AGN luminosity function can also be expressed as
\begin{equation}
\frac{d\phi}{dL} = \frac{2.5\phi_{\star}/\ln{10}}{(L/L_{\star})^{-\alpha} + (L/L_{\star})^{-\beta}},
\end{equation}
where $L_{\star}$ is the break luminosity corresponding to $M_{\star}$.

Since the emissivity $\epsilon = \int dL L (d\phi/dL)=\int d(\ln{L}) L^{2} (d\phi/dL)$, each logarithmic bin of luminosity contributes $d(\ln{L}) L^{2} (d\phi/dL)$ to the total emissivity. 
Defining $x=L/L_{\star}$ and expressing in terms of luminosity decades, the double power-law AGN luminosity function implies a fractional contribution to the total emissivity per luminosity decade
\begin{align}
\frac{1}{\epsilon} \frac{d\epsilon}{d\log_{10}{L}} =
 \frac{ x^{2}/(x^{-\alpha}+x^{-\beta})}
{ \int_{0}^{\infty} dx' x'/(x'^{-\alpha}+x'^{-\beta}) }.
\end{align}
Figure \ref{fig:QLF_completeness_diff} shows how different AGN luminosity bins contribute to the total emissivity. 
On the left, we show the results for double power-law models with varying faint-end slopes while on the right we show the results for the Hopkins et al. (2007), Kulkarni et al. (2019), and \cite{2020arXiv200102696S} luminosity functions at $z=0.25$, as a function of UV luminosity at rest-frame 1450~\AA. 
We note that the Hopkins et al. and Shen et al. UV luminosity functions do not follow exact double power-law models, since they are computed from (observationally calibrated) models for the AGN bolometric luminosity function, taking into account luminosity-dependent obscuration and scatter in the bolometric corrections. 
The figure shows that the Shen et al. luminosity function used in this work converges more rapidly with decreasing limiting luminosity than the Kulkarni et al. luminosity function.

\section{Photoionization/heating rates}
\label{sec:rates_appendix}
Table \ref{tbl:rates} lists the HI, HeI, and HeIII photoionization and photoheating rates for our UVB model and the fiducial reionization history parameters in \S \ref{sec:reionization_params}. 
Before reionization events, these rates are effective homogeneous values appropriate for use in simulation codes that assume photoionization equilibrium with a homogenous UVB.

\begin{footnotesize}
\ctable[
  caption={{\normalsize Effective Photoionization/heating Rates for the Fiducial Planck 2018 Reionization History}\label{tbl:rates}},center,star
  ]{lcccccccl}{
\tnote[ ]{Reionization history parameters (defined in \S \ref{sec:reionization}). 
HI reionization: $z_{\rm rei,HI} = 7.8$, $\Delta z_{\rm rei,HI} = 0.5$, $C_{\rm HI}=3$, $T_{\rm IGM0,HI}=10,000$ K, $\Delta T_{\rm HI}=20,000$ K. 
HeII reionization: $z_{\rm rei,HeII} = 3.5$, $\Delta z_{\rm rei,HeII} = 0.5$, $C_{\rm HeII}=3$, $T_{\rm IGM0,HeII}=10,000$ K, $\Delta T_{\rm HeII}=15,000$ K. 
}
}{
\hline\hline
\multicolumn{1}{c}{$z$} &
\multicolumn{1}{c}{$\Gamma_{\rm HI}^{\rm eff}$} &
\multicolumn{1}{c}{$\Gamma_{\rm HeI}^{\rm eff}$} & 
\multicolumn{1}{c}{$\Gamma_{\rm HeII}^{\rm eff}$} & 
\multicolumn{1}{c}{$\dot{q}_{\rm HI}^{\rm eff}$} & 
\multicolumn{1}{c}{$\dot{q}_{\rm HeI}^{\rm eff}$} & 
\multicolumn{1}{c}{$\dot{q}_{\rm HeII}^{\rm eff}$} \\ 
\multicolumn{1}{c}{\ } &
\multicolumn{1}{c}{10$^{-12}$ s$^{-1}$} &
\multicolumn{1}{c}{10$^{-12}$ s$^{-1}$} &
\multicolumn{1}{c}{10$^{-12}$ s$^{-1}$} &
\multicolumn{1}{c}{10$^{-24}$ erg s$^{-1}$} &
\multicolumn{1}{c}{10$^{-24}$ erg s$^{-1}$} &
\multicolumn{1}{c}{10$^{-24}$ erg s$^{-1}$} \\ 
\hline
$0.0$ & $3.62\times 10^{-2}$ & $1.19\times 10^{-2}$ & $2.84\times 10^{-4}$ & $0.203$ & $0.127$ & $8.77\times 10^{-3}$ \\ 
$0.2$ & $8.72\times 10^{-2}$ & $3.50\times 10^{-2}$ & $6.71\times 10^{-4}$ & $0.516$ & $0.359$ & $1.93\times 10^{-2}$ \\ 
$0.4$ & $0.177$ & $8.04\times 10^{-2}$ & $1.40\times 10^{-3}$ & $1.075$ & $0.837$ & $3.81\times 10^{-2}$ \\ 
$0.6$ & $0.312$ & $0.152$ & $2.58\times 10^{-3}$ & $1.926$ & $1.635$ & $6.72\times 10^{-2}$ \\ 
$0.8$ & $0.488$ & $0.246$ & $4.20\times 10^{-3}$ & $3.034$ & $2.732$ & $0.105$ \\ 
$1.0$ & $0.682$ & $0.352$ & $6.00\times 10^{-3}$ & $4.260$ & $3.995$ & $0.146$ \\ 
$1.2$ & $0.862$ & $0.452$ & $7.64\times 10^{-3}$ & $5.414$ & $5.218$ & $0.182$ \\ 
$1.4$ & $1.003$ & $0.533$ & $8.74\times 10^{-3}$ & $6.338$ & $6.229$ & $0.203$ \\ 
$1.6$ & $1.091$ & $0.590$ & $9.12\times 10^{-3}$ & $6.939$ & $6.950$ & $0.207$ \\ 
$1.8$ & $1.136$ & $0.622$ & $9.10\times 10^{-3}$ & $7.264$ & $7.386$ & $0.198$ \\ 
$2.0$ & $1.136$ & $0.631$ & $8.58\times 10^{-3}$ & $7.316$ & $7.518$ & $0.180$ \\ 
$2.2$ & $1.109$ & $0.627$ & $7.47\times 10^{-3}$ & $7.191$ & $7.432$ & $0.153$ \\ 
$2.4$ & $1.069$ & $0.611$ & $6.23\times 10^{-3}$ & $6.960$ & $7.196$ & $0.125$ \\ 
$2.6$ & $1.021$ & $0.589$ & $5.02\times 10^{-3}$ & $6.678$ & $6.874$ & $9.76\times 10^{-2}$ \\ 
$2.8$ & $0.968$ & $0.564$ & $3.91\times 10^{-3}$ & $6.359$ & $6.508$ & $7.33\times 10^{-2}$ \\ 
$3.0$ & $0.915$ & $0.537$ & $2.14\times 10^{-3}$ & $6.027$ & $6.112$ & $5.19\times 10^{-2}$ \\ 
$3.2$ & $0.857$ & $0.509$ & $2.30\times 10^{-4}$ & $5.671$ & $5.716$ & $1.65\times 10^{-2}$ \\ 
$3.4$ & $0.805$ & $0.482$ & $8.23\times 10^{-5}$ & $5.349$ & $5.340$ & $8.34\times 10^{-3}$ \\ 
$3.6$ & $0.757$ & $0.457$ & $3.16\times 10^{-5}$ & $5.035$ & $4.995$ & $4.91\times 10^{-3}$ \\ 
$3.8$ & $0.711$ & $0.431$ & $8.08\times 10^{-6}$ & $4.744$ & $4.679$ & $2.42\times 10^{-3}$ \\ 
$4.0$ & $0.670$ & $0.410$ & $2.81\times 10^{-7}$ & $4.465$ & $4.393$ & $2.80\times 10^{-4}$ \\ 
$4.2$ & $0.631$ & $0.388$ & $1.00\times 10^{-28}$ & $4.203$ & $4.136$ & $1.00\times 10^{-16}$ \\ 
$4.4$ & $0.592$ & $0.365$ & $1.00\times 10^{-28}$ & $3.946$ & $3.899$ & $1.00\times 10^{-16}$ \\ 
$4.6$ & $0.558$ & $0.348$ & $1.00\times 10^{-28}$ & $3.718$ & $3.706$ & $1.00\times 10^{-16}$ \\ 
$4.8$ & $0.525$ & $0.331$ & $1.00\times 10^{-28}$ & $3.502$ & $3.526$ & $1.00\times 10^{-16}$ \\ 
$5.0$ & $0.493$ & $0.315$ & $1.00\times 10^{-28}$ & $3.300$ & $3.369$ & $1.00\times 10^{-16}$ \\ 
$5.2$ & $0.463$ & $0.301$ & $1.00\times 10^{-28}$ & $3.116$ & $3.233$ & $1.00\times 10^{-16}$ \\ 
$5.4$ & $0.436$ & $0.287$ & $1.00\times 10^{-28}$ & $2.947$ & $3.101$ & $1.00\times 10^{-16}$ \\ 
$5.6$ & $0.409$ & $0.273$ & $1.00\times 10^{-28}$ & $2.779$ & $2.971$ & $1.00\times 10^{-16}$ \\ 
$5.8$ & $0.384$ & $0.259$ & $1.00\times 10^{-28}$ & $2.615$ & $2.835$ & $1.00\times 10^{-16}$ \\ 
$6.0$ & $0.356$ & $0.245$ & $1.00\times 10^{-28}$ & $2.441$ & $2.704$ & $1.00\times 10^{-16}$ \\ 
$6.2$ & $0.329$ & $0.230$ & $1.00\times 10^{-28}$ & $2.268$ & $2.564$ & $1.00\times 10^{-16}$ \\ 
$6.4$ & $0.304$ & $0.216$ & $1.00\times 10^{-28}$ & $2.102$ & $2.418$ & $1.00\times 10^{-16}$ \\ 
$6.6$ & $0.279$ & $0.202$ & $1.00\times 10^{-28}$ & $1.943$ & $2.282$ & $1.00\times 10^{-16}$ \\ 
$6.8$ & $0.257$ & $0.189$ & $1.00\times 10^{-28}$ & $1.801$ & $2.160$ & $1.00\times 10^{-16}$ \\ 
$7.0$ & $0.237$ & $0.177$ & $1.00\times 10^{-28}$ & $1.670$ & $2.040$ & $1.00\times 10^{-16}$ \\ 
$7.2$ & $0.167$ & $0.125$ & $1.00\times 10^{-28}$ & $1.188$ & $1.446$ & $1.00\times 10^{-16}$ \\ 
$7.4$ & $1.25\times 10^{-2}$ & $8.03\times 10^{-3}$ & $1.00\times 10^{-28}$ & $0.107$ & $9.42\times 10^{-2}$ & $1.00\times 10^{-16}$ \\ 
$7.6$ & $1.44\times 10^{-4}$ & $9.96\times 10^{-6}$ & $1.00\times 10^{-28}$ & $5.77\times 10^{-3}$ & $4.81\times 10^{-4}$ & $1.00\times 10^{-16}$ \\ 
$7.8$ & $2.80\times 10^{-5}$ & $1.93\times 10^{-6}$ & $1.00\times 10^{-28}$ & $2.43\times 10^{-3}$ & $2.03\times 10^{-4}$ & $1.00\times 10^{-16}$ \\ 
$8.0$ & $4.26\times 10^{-6}$ & $2.94\times 10^{-7}$ & $1.00\times 10^{-28}$ & $1.11\times 10^{-3}$ & $9.27\times 10^{-5}$ & $1.00\times 10^{-16}$ \\ 
$8.2$ & $2.81\times 10^{-7}$ & $1.94\times 10^{-8}$ & $1.00\times 10^{-28}$ & $3.56\times 10^{-4}$ & $2.97\times 10^{-5}$ & $1.00\times 10^{-16}$ \\ 
$8.4$ & $1.24\times 10^{-9}$ & $8.62\times 10^{-11}$ & $1.00\times 10^{-28}$ & $3.13\times 10^{-5}$ & $2.61\times 10^{-6}$ & $1.00\times 10^{-16}$ \\ 
$8.6$ & $1.00\times 10^{-28}$ & $1.00\times 10^{-28}$ & $1.00\times 10^{-28}$ & $1.00\times 10^{-16}$ & $1.00\times 10^{-16}$ & $1.00\times 10^{-16}$ \\ 
$8.8$ & $1.00\times 10^{-28}$ & $1.00\times 10^{-28}$ & $1.00\times 10^{-28}$ & $1.00\times 10^{-16}$ & $1.00\times 10^{-16}$ & $1.00\times 10^{-16}$ \\ 
\hline\hline\\
}
\end{footnotesize}

\bibliography{references}

\end{document}